\documentclass[a4paper,fleqn,usenatbib]{mnras}

\usepackage[T1]{fontenc}
\usepackage{ae,aecompl}

\usepackage{graphicx}	
\usepackage{amsmath}	
\usepackage{amssymb}	

\title[Photometric and polarimetric variability of 3C~323.1]{Constraints on the optical polarization source in the luminous non-blazar quasar 3C 323.1 (PG 1545+210) from the photometric and polarimetric variability}

\author[M.~Kokubo]{
Mitsuru Kokubo,$^{1,2}$\thanks{E-mail: mkokubo@ioa.s.u-tokyo.ac.jp}
\\
$^{1}$Department of Astronomy, School of Science, the University of Tokyo, 7-3-1 Hongo, Bunkyo-ku, Tokyo 113-0033, Japan\\
$^{2}$Institute of Astronomy, the University of Tokyo, 2-21-1 Osawa, Mitaka, Tokyo 181-0015, Japan\\
}

\date{Accepted 2017 January 10. Received 2017 January 08; in original form 2016 November 29}

\pubyear{2017}

\begin{document}
\label{firstpage}
\pagerange{\pageref{firstpage}--\pageref{lastpage}}
\maketitle

\begin{abstract}

We examine the optical photometric and polarimetric variability of the luminous type~1 non-blazar quasar 3C~323.1 (PG 1545+210). 
Two optical spectro-polarimetric measurements taken during the periods 1996$-$98 and 2003 combined with a $V$-band imaging polarimetric measurement taken in 2002 reveal that 
(1) as noted in the literature, the polarization of 3C 323.1 is confined only to the continuum emission, that is, the emission from the broad line region is unpolarized; 
(2) the polarized flux spectra show evidence of a time-variable broad absorption feature in the wavelength range of the Balmer continuum and other recombination lines; 
(3) weak variability in the polarization position angle ($PA$) of $\sim$ 4 deg over a time-scale of 4$-$6 years is observed; and 
(4) the V-band total flux and the polarized flux show highly correlated variability over a time-scale of one year. Taking the above-mentioned photometric and polarimetric variability properties and the results from previous studies into consideration, we propose a geometrical model for the polarization source in 3C~323.1, in which an equatorial absorbing region and an axi-asymmetric equatorial electron-scattering region are assumed to be located between the accretion disc and the broad line region. The scattering/absorbing regions can perhaps be attributed to the accretion disc wind or flared disc surface, but further polarimetric monitoring observations for 3C~323.1 and other quasars with continuum-confined polarization are needed to probe the true physical origins of
these regions.

\end{abstract}





\section{Introduction}
\label{sec:intro}

It has long been known that the ultraviolet(UV)-optical emission of non-blazar active galactic nuclei (AGNs) often shows weak linear polarization.
The polarization observed in type 2 AGNs is considered to be the result of electron/dust scattering of the AGN nuclear emission from the polar scattering region; photons from the accretion disc and the broad line region (BLR) behind the obscuring dust torus are scattered into the observer's line of sight by the polar scattering region and, thus, the blue disc continuum and BLR emission appear in the polarized flux spectra \citep{ant85}.
Observations have revealed that the polarization position angle ($PA$) in type 2 objects is generally perpendicular to the radio jet structure (note that the jet axis is thought to be parallel to the accretion disc's rotation axis), as expected by the polar scattering model \citep[e.g.,][]{ant83,bri90,smi04}.
On the other hand, in the case of type 1 AGNs with a prominent radio jet structure, polarization is generally known to be shown parallel to the radio structure \citep[e.g.,][]{sto79,ant83,sto84,sch00,smi02,kis04}.
Even for radio-quiet objects, high-spatial-resolution imaging observations have revealed there to be a significant correlation between the direction of the extended emission region (which is another good tracer of the accretion disc's rotation axis) and the polarization PA, in that the former tends to be parallel to the latter in type 1 objects and perpendicular in type 2 objects \citep{zak05,bor08}.
It is generally difficult to explain the ``parallel'' polarization observed in type 1 objects by the polar scattering or the polarization induced within the atmosphere of a plane-parallel scattering-dominated disc \citep[e.g.,][]{ant88,kis03}.
In addition, some authors have claimed that the intrinsic quasar accretion disc continua must be completely depolarized because of the strong Faraday depolarization with magnetic fields in the disc atmosphere \citep[][and references therein]{ago96,sil09}.
Instead, the observed optical polarization properties in type 1 quasars can be understood by assuming a geometrically and optically thin equatorial electron-scattering region located inside the dust torus; photons produced inside this electron-scattering region can be scattered into our line of sight, resulting in a net linear polarization ``parallel'' to the disc rotation axis  \citep[e.g.,][]{sto79,ant88,smi04,smi05,goo07,kis08b,bat11,gas12,mar13,hut15,sil16}.
It is currently believed that all AGNs/quasars have both equatorial and polar scattering regions, and that differences in the observed polarization properties between type 1 and type 2 objects are due to the orientation effect \citep[e.g.,][]{smi04}.

In most type~1 Seyfert galaxies, it is observationally known that not only are the UV-optical continua polarized but also the BLR emission lines.
polarization of the BLR emission lines is often observed to be at a lower polarization degree and a different $PA$ than the continuum emission, and the rotation of the $PA$ can be seen as a function of the wavelength \citep[e.g.,][]{smi02,smi04}
This implies that the equatorial electron scattering region is more or less similar in size to the BLR in these AGNs \citep[e.g.,][]{ang80,smi05,kis08b,bal16}.
However, optical spectro-polarimetric observations for luminous type 1 quasars\footnote{In this paper, we use the term ``quasars'' to refer to optically identified bright AGNs, including the radio-loud and the radio-quiet objects.} with ``parallel'' optical polarization carried out by \cite{ant88}, \cite{sch00} and \cite{kis03,kis04,kis05,kis08} have revealed that there is a population of quasars whose BLR emission lines are essentially unpolarized, i.e., the polarization of these quasars is confined only to the continuum.
The observed null polarization of the BLR emission rules out the possibility that the optical polarization in these quasars is predominantly attributed to scattering processes outside the BLR \citep[e.g.,][]{smi93,kis03}.
The continuum polarization PAs of these quasars are mostly wavelength-independent, indicating that there is a single dominant mechanism producing the observed polarization.

\cite{kis03,kis04} interpreted the continuum-confined polarization observed in several type 1 quasars such that the equatorial electron scattering region in these quasars is smaller in size compared to the BLR, but still much larger than the UV-optical emitting regions of the accretion disc.
In such a geometrical configuration, electron scattering produces no net polarization in the observed BLR emission because the polarization is cancelled out when the polarized BLR flux vectors are averaged over a wide range of angles, and thus the polarization due to equatorial electron scattering is confined only to the accretion disc continua \citep[see also][]{kis08b}.
One interesting consequence of Kishimoto's interpretation is that because the electron (Thomson) scattering cross-section is wavelength-independent, the polarized flux spectra in these quasars should reflect the spectral shape of the intrinsic accretion disc continua hidden under the strong BLR emission \citep{ant02,kis03,kis04,hu12,mar13}.
Interestingly, \cite{kis04} showed that the polarized flux spectra of quasars with continuum-confined polarization share a similar spectral break feature at $\lambda_{\rm{rest}}\sim3600$\AA, which they interpreted as the Doppler-broadened hydrogen Balmer absorption edge imprinted in the intrinsic accretion disc thermal emission due to the disc atmospheric opacity effect.
\cite{kis04} further claimed that the overall spectral shape of these quasars (including the Balmer edge-like spectral break) can be well described by the non-local thermodynamic equilibrium radiative-transferred thermal accretion disc model of \cite{hub00} \citep[see also][]{hu12}.

However, \cite{kok16} has pointed out that the spectral shape of the accretion disc continua of quasars with continuum-confined polarization revealed as the variable component spectra (see Section~\ref{obs_properties}) contradicts that revealed as the polarized component spectra, which may imply that equatorial electron scattering of the disc continua alone probably cannot explain the optical polarization properties in these quasars.
Moreover, it should be noted that, before the deep spectro-polarimetric observations conducted by \cite{kis04}, \cite{sch00} had carried out spectro-polarimetry for quasar samples overlapping with those used by \cite{kis04} and suggested that the polarization in quasars with continuum-confined polarization may be attributed to a weak optical synchrotron component from the misdirected blazar core components.
In fact, all of the five quasars (3C~323.1, Ton~202, B2 1208+32, 3C~95, and 4C09.72) confirmed as showing continuum-confined polarization by both \cite{sch00} and \cite{kis04} are radio-loud objects; thus, it is unsurprising that there is a flux contribution from the jet synchrotron component to the observed optical spectra (see e.g., \citealt{imp89,smi93} and \citealt{afa15a} for the cases of 3C~273 and 3C~390.3).
However, it should be noted that subsequent work by \cite{kis08} has identified two radio-quiet quasars (Q0144-3938 and CTS~A09.36) with continuum-confined polarization.

As discussed above, the optical polarization source in quasars with continuum-confined polarization has yet to be fully understood.
To probe the true nature of the optical polarization mechanism, and consequently to examine the accretion disc physics in luminous type 1 quasars, detailed studies of the quasar polarimetric variability must be a key observable to constrain the geometry of the polarization source \citep[e.g.,][]{sto79,sch00,gal05,gas12,afa15a,roj16}.
In this work we examine the variability of the total flux, polarization degree, polarization PA, and the polarized flux in the luminous type 1 non-blazar quasar 3C~323.1, which is a quasar with continuum confined polarization identified by \cite{sch00} and \cite{kis04}, by collecting archival data.

The three polarimetry data sets used in this work, including the two optical spectro-polarimetric measurements taken by \cite{sch00} (Bok/SPOL in 1996-1998) and by \cite{kis04} (Keck/LRIS in 2003) and a $V$-band imaging-polarimetry measurement by \cite{slu05} (ESO3.6-m/EFOSC2 in 2002), are described in Section~\ref{data_3c3231}.
The absolute $V$-band magnitudes at the epochs of the spectro-polarimetric measurements are carefully estimated by using broad-band light curves taken from the literature.
In Section~\ref{spec_var_properties}, we first examine the spectral variability of the polarimetric properties between the two spectro-polarimetric measurements, and identify the origin of the different conclusions between \cite{sch00} and \cite{kis04} regarding the polarization source in the quasars with continuum-confined polarization described above.
We then examine the $V$-band photometric and polarimetric variability of 3C~323.1 using all three polarimetric measurements in order to constrain observationally the geometry of the optical polarization source in 3C~323.1 in Section~\ref{discuss}.
Finally, a summary and conclusions are provided in Section~\ref{summary_and_conclusion_3c3231}.

\section{Data}
\label{data_3c3231}

Throughout this paper, we assume a cosmology of $H_0=73$ km/s/Mpc, $\Omega_{m}=0.27$, and $\Omega_{\Lambda}=0.73$ \citep{spe07}.
None of the data are corrected for Galactic extinction.
In this study, ``the total flux'' ($F_{\lambda}$ or $F_{\nu}$) has the same meaning as the un-normalised Stokes parameter $I$ expressed in units of the flux.
$q$ and $u$ represent the normalised Stokes parameters defined as $Q/I$ and $U/I$, whose polarization $PA$ is defined using the equatorial coordinate system (a $PA$ of 0 deg corresponds to North-South, and the $PA$ increases from North towards East).
The polarization degree $p\equiv\sqrt{q^2+u^2}$ obtained from the observational data with finite measurement errors\footnote{For the polarimetric observations for weakly polarized objects, the measurement errors associated with $q$ and $u$ is essentially the same ($\sigma\equiv \sigma_q \simeq \sigma_u$).} is known to be a biased estimate of the true polarization degree; therefore, the de-biased estimate of the polarization degree, $\hat{p}\equiv \sqrt{p^2-\sigma^2}$ \citep[e.g.,][]{pla14}, is used throughout this study.
The rotated normalised Stokes parameters $q'$ and $u'$ are defined in a coordinate system rotated to the systemic polarization position angle $PA_{c}$ of the target; note that all of the polarization, in principle, falls in the single Stokes parameter $q'$ when the object has a wavelength-independent $PA$.

\subsection{Observational properties of 3C~323.1: the variable component spectrum}
\label{obs_properties}

3C~323.1 (PG~1545+210) is a Fanaroff-Riley~II (FRII) radio-loud quasar at $z=0.264$ \citep[the luminosity distance $d_L=1293.3$ Mpc;][]{wri06}.
The lobe-dominant radio jet structure of 3C~323.1 has a size of $\sim$ 300 kpc, and its position angle is $\sim$ 20 deg \citep{mil78,kel94,den00,sch00,kis04}.
As is the case with the other quasars with the continuum-confined polarization, the radio jet axis of 3C~323.1 is known to be parallel to the direction of the optical polarization vector within $\lesssim$ 10 deg \citep[see Section~\ref{discuss};][]{sto79,sch00,kis04}.
In the optical wavelength range, 3C~323.1 is observed as a point source with an apparent $V$-band magnitude of $V\sim16$ mag.
The optical multi-band colour of 3C~323.1 is within the spectroscopic target selection criteria in the Sloan Digital Sky Survey (SDSS) Legacy Survey \citep{ric02}, meaning that this object is ``normal'' in terms of its optical spectrum \citep[see also][]{bor92}.
The black hole mass ($M_{BH}$) and the Eddington ratio (defined as the ratio of the bolometric luminosity $L$ to the Eddington luminosity $L_{E}$) of 3C~323.1 were estimated to be log($M_{\text{BH}}$/$M_{\odot})=9.07$ and $L/L_{E}=0.10$, respectively, based on the single-epoch SDSS spectrum obtained at MJD=53886 \citep{she11}.

Recently, \cite{kok16} has presented the multi-band ($u$, $g$, $r$, $i$, and $z$-band) photometric light curves for 3C~323.1 obtained in 2015-2016, and has examined the spectral shape of the UV-optical variable component spectrum in this object.
\cite{kok16} showed that the variable component spectrum in 3C~323.1 is consistent with a power-law spectrum with $\alpha_{\nu}\sim+1/3$ (in the form of $F_{\nu}\propto\nu^{\alpha_{\nu}}$).
The spectral index of $\alpha_{\nu}\sim+1/3$ is commonly detected in the variable component spectra of (mostly radio-quiet) quasars \citep[e.g.,][]{per06,rua14,kok14,mac16,hun16,bui17}, and is usually attributed to the well-known prediction of the spectral index $\alpha_{\nu}=+1/3$ of the standard thin thermal accretion disc emission \citep{sha73,nov73,fra92}.
This means that it is natural to consider that the flux variability of 3C~323.1 is due to the intrinsic variations of the disc emission itself, in line with other non-blazar quasars; nevertheless, the precise mechanisms for the quasar flux variability are still under debate \citep[see][and references therein]{kok15,kok16}.
The observed variability time-scale of months to years in 3C~323.1 is also consistent with the variability of the quasar accretion disc emission.
As discussed in Section~\ref{discuss}, the attribution of the flux variability to the intrinsic disc emission variability, combined with the observed correlated variability between the total flux and the polarized flux, implies that the polarized flux is directly related to the disc emission and does not arise from the optical synchrotron flux contribution.

\subsection{Polarimetry data}
\label{pol_data_description}

\begin{figure}
\center{
\includegraphics[clip, width=2.7in]{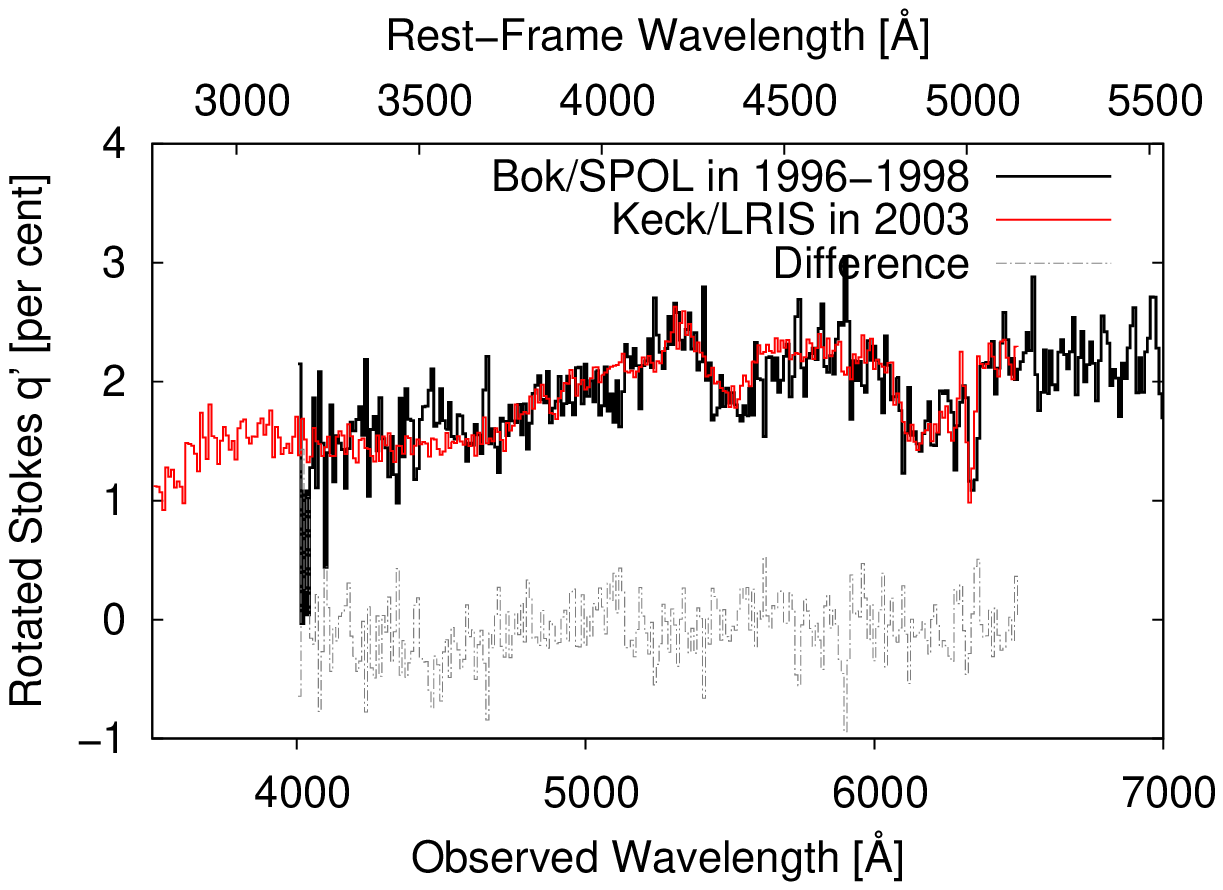}
\includegraphics[clip, width=2.7in]{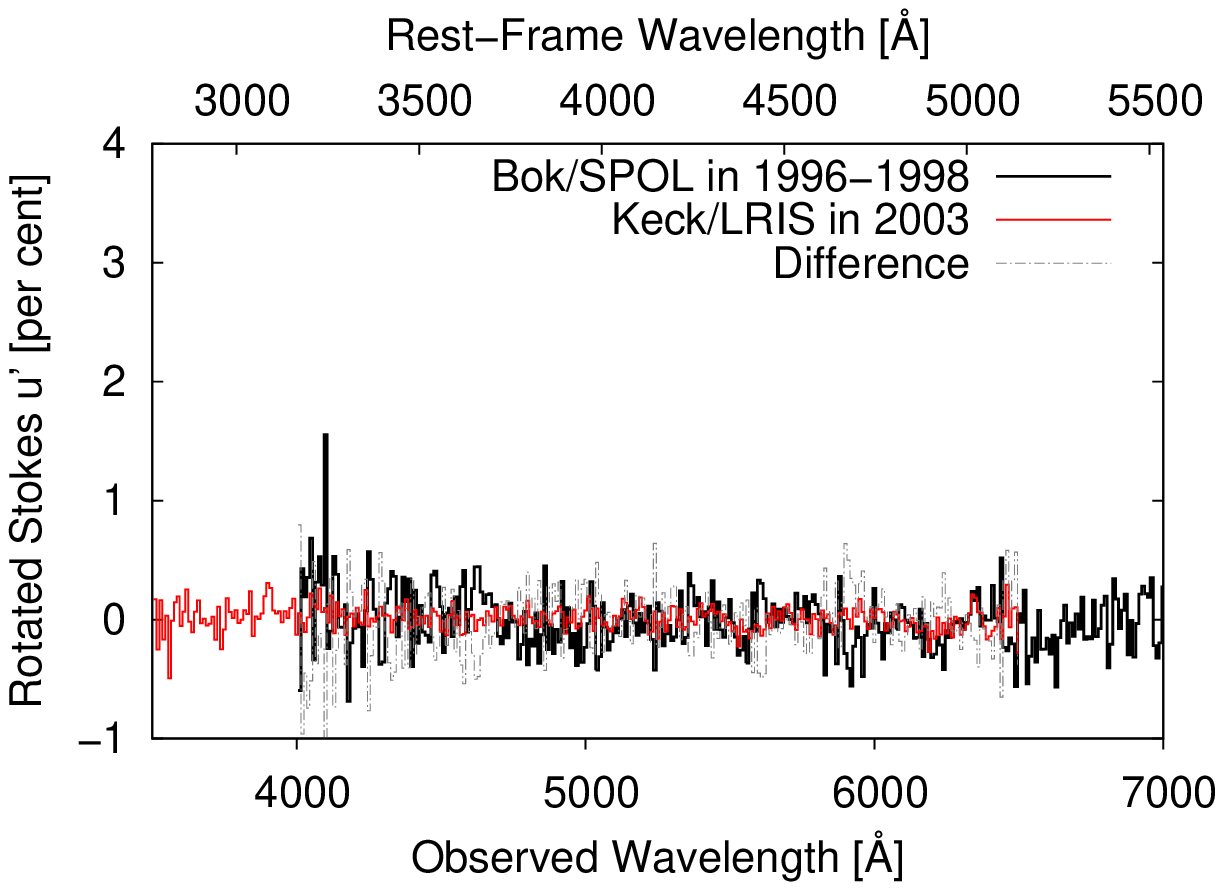}
\includegraphics[clip, width=2.7in]{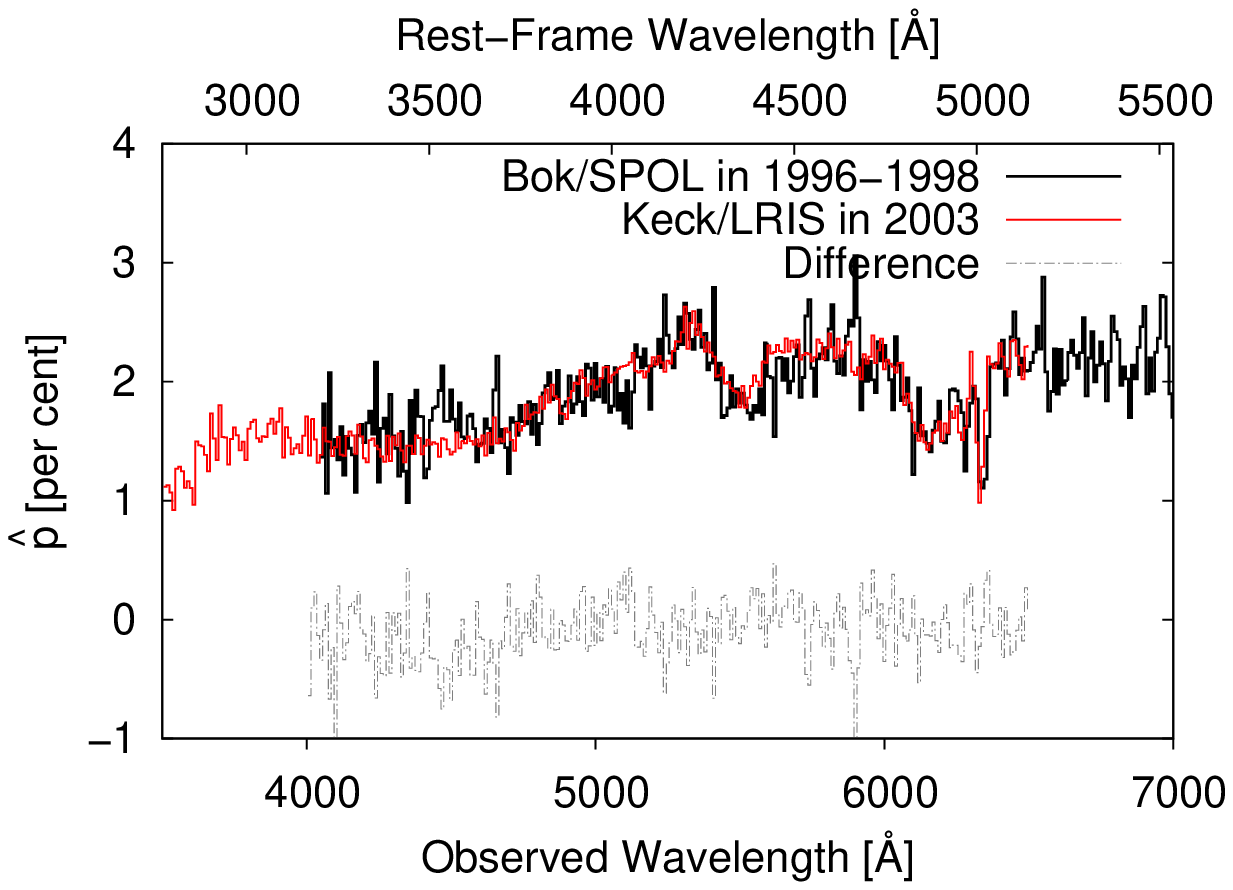}
\includegraphics[clip, width=2.7in]{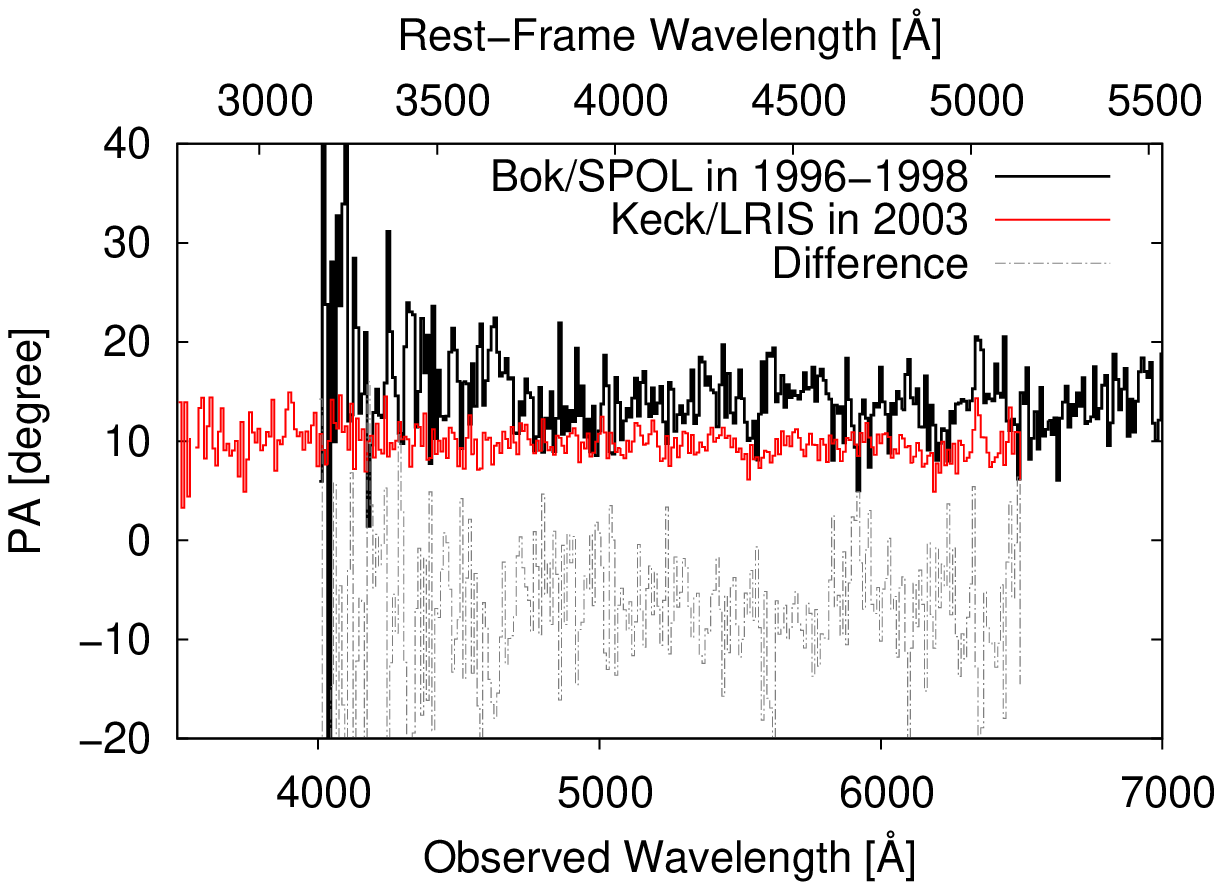}
}
 \caption{Comparison of the rotated Stokes parameters $q'$ and $u'$, de-biased polarization degree $\hat{p}$, and the polarization position angle ($PA$) of 3C~323.1 between the spectro-polarimetric measurements of Bok/SPOL \citep{sch00} and Keck/LRIS \citep{kis04}.
 The difference spectra between the two measurements are also plotted.
The spectra are binned into common wavelength bins with 10 \AA\ width to allow easy comparison between the two measurements.
Note that $q'$ and $u'$ are defined by assuming the systemic $PA$ as $PA_{c}=$14.2 deg and 9.6 deg for Bok/SPOL and Keck/LRIS data, respectively.
}
 \label{fig:fig_pol}
\end{figure}

In this work, we use the three archival polarimetric measurements for 3C~323.1 obtained at different epochs with Bok/SPOL, ESO3.6-m/EFOSC2, and Keck/LRIS.
The ESO3.6-m/EFOSC2 observation is $V$-band imaging polarimetry \citep{slu05}, and the Bok/SPOL and the Keck/LRIS observations are optical spectro-polarimetry \citep{sch00,kis04}.
The compilation of these polarimetric data from the literature enables to examine the two-epoch spectral variability (Section~\ref{spec_var_properties}) and the three-epoch $V$-band variability (Section~\ref{discuss}) of the polarimetric properties of 3C~323.1.

The $V$-band magnitudes of the total flux at the epochs of the three polarimetric measurements are also estimated  (Sections~\ref{eso_imapol} and ~\ref{spectrophotometry}).
As discussed in Section~\ref{discuss}, one of the most important subjects of this work is to examine the correlation between the total flux and the polarized flux light curves, which eventually provides strong constraints on the optical polarization source in 3C~323.1.

Following the same procedure of \cite{kis04}, we correct the three polarimetric measurements for the effects of Galactic interstellar polarization (ISP).
The details of the method of the ISP correction is summarized in Section~\ref{appendix_isp_correction}.
Throughout this paper, the characters with no subscript represent ISP-corrected values unless otherwise stated.

\subsubsection{ESO3.6-m/EFOSC2 $V$-band imaging-polarimetry}
\label{eso_imapol}

\cite{slu05} presented the results of $V$-band imaging-polarimetry for 3C~323.1 obtained from the 3.6-m telescope at the European Southern Observatory (ESO), La Silla, equipped with the ESO Faint Object Spectrograph and Camera (v.2) (EFOSC2) on the 1st May, 2002 \citep[MJD=52396.3; see also][]{hut05}.
As part of this, they took a set of four 60-sec integration images with the half-wave plate position angles of 0.0 deg, 22.5 deg, 45.0 deg, and 67.5 deg.

We downloaded the ESO3.6-m/EFOSC2 imaging-polarimetry data of 3C~323.1 and the associated {\tt BIAS} and {\tt FlatPolIma} frames from the ESO Science Archive Facility (Program ID 68.A-0373).
After applying bias and flat corrections, the $V$-band polarization degree $\hat{p}_V$ and the polarization position angle $PA_V$ of 3C~323.1 were calculated by adopting the aperture photometry and the ``ratio'' method \citep[e.g.,][]{lam99,bag09}.
By subtracting the instrumental polarization using the values presented in \cite{slu05}, we obtained the ISP-uncorrected $V$-band polarization degree and polarization position angle as $p_{V, \text{uncorr}}=1.14\pm0.13$ \%, $\hat{p}_{V, \text{uncorr}}=1.13$ \% and $PA_{V, \text{uncorr}}=17.88\pm3.39$ deg, which are consistent with those reported by \cite{slu05}\footnote{\cite{slu05} reported the ISP-uncorrected $V$-band polarization degree and the polarization position angle of 3C~323.1 as $p_{V, \text{uncorr}}=1.15\pm0.13$\%, $\hat{p}_{V, \text{uncorr}}=1.14$\% and $PA_{V, \text{uncorr}}$=18$\pm$3 deg. }.
By correcting for the Galactic ISP (Section~\ref{appendix_isp_correction}), we obtain $\hat{p}_{V}=1.94$ \% and $PA_{V}=10.03$ deg (Table~\ref{obs_Vband}).

To evaluate the $V$-band magnitude at the epoch of the EFOSC2 polarimetry measurement, we applied differential photometry by using two field stars (SDSS J154736.44+205141.7 and SDSS J154752.41+204931.8) simultaneously imaged on the same EFOSC2 frames along with 3C~323.1 as reference stars.
By using the SDSS $g$ and $r$-band point spread function (PSF) photometry and the transformation equation by \cite{jes05} from the SDSS $g$ and $r$-band magnitude to $V$-band magnitude, $V_{Vega}=g-0.59(g-r)-0.01$ mag with an uncertainty of 0.01 mag, the $V$-band magnitudes of SDSS J154736.44+205141.7 and SDSS J154752.41+204931.8 are estimated as $V_{Vega}=17.0516$ mag and $V_{Vega}=16.6971$ mag, respectively.
For each image, aperture photometry was applied to combine the two orthogonally polarized fluxes of 3C~323.1 and the two reference stars.
By adding the 0.01 mag error due to the transformation equation of \cite{jes05} to the photometric errors, we evaluated the weighted average magnitude of 3C~323.1 and its uncertainty as $V_{Vega}=15.7779\pm0.0106$ mag or $V_{AB}=15.7339\pm0.0113$ mag, where the AB offset and its uncertainty are taken from \cite{fre94}.
The polarimetric and photometric measurements for 3C~323.1 obtained with ESO3.6-m/EFOSC2 are summarised in Table~\ref{obs_Vband}.

\subsubsection{Bok/SPOL and Keck/LRIS optical spectro-polarimetry}
\label{bol_keck_specpol}

\begin{table*}
	\centering
	\caption{Summary of the ISP-corrected $V$-band polarimetric and photometric values for 3C~323.1.}
	\label{obs_Vband}
	\begin{tabular}{lllccccl}
		\hline
		Date & MJD & Instrument & $\hat{p}_{V}$[\%] & $PA_{V}$[deg] & V-band AB mag. of the total flux\\
		\hline
1996-06 and 1998-04&50235-50933 & Bok/SPOL & 2.06$\pm$0.01& 13.75$\pm$0.20&$15.8935\pm 0.0358$\\
2002-05-01&52396.3  & ESO3.6-m/EFOSC2 & 1.94$\pm$0.13 & 10.03$\pm$3.39&$15.7339\pm 0.0113$ \\
2003-05-04&52763.6  & Keck/LRIS & 2.11$\pm$0.01 & 9.61$\pm$0.10&$15.5838\pm 0.0226$\\
		\hline
	\end{tabular}
\end{table*}

3C~323.1 was spectro-polarimetrically observed by \cite{sch00} and \cite{kis04} during 1996-1998 and 2003, respectively.

\cite{sch00} carried out spectro-polarimetric observations of 3C~323.1 at the Steward Observatory 2.3-m Bok Telescope using the SPOL CCD Imaging/Spectropolarimeter \citep[][]{sch92} with a low-resolution grating.
The observed wavelength range was $\lambda_{obs}=4000-8000$ \AA.
The data presented in \cite{sch00} comprise 10 observations from several epochs taken in June 1996 (5 observations) and April 1998 (5 observations), and thus the combined data represent the average polarization properties of 3C~323.1 at MJD$\sim$50235 and 50933.
Since the raw Bok/SPOL data of 3C~323.1 are unavailable, in this work, we directly use the combined calibrated polarization spectra of 3C~323.1 presented in \cite{sch00}, which are kindly provided by G.~D.~Schmidt and P.~S.~Smith in electric form.
The Galactic ISP is corrected as described in Section~\ref{appendix_isp_correction}.
Following the definition of \cite{kis04}, the systemic polarization position angle $PA_{c}$ of 3C~323.1 at the epoch of Bok/SPOL observations is evaluated to be $PA_{c} = 14.2$ deg as a weighted average of the $PA$ at the rest-frame wavelength range of 4000-4731\AA; thus, the rotated Stoke parameters $q'$ and $u'$ are defined in a coordinate system rotated to $PA_{c} = 14.2$ deg.

\cite{kis04} carried out spectro-polarimetry for 3C~323.1 on the 4th May, 2003 (MJD=52763.6), with the Low Resolution Imaging Spectrometer (LRIS) mounted on the 10-m Keck-I telescope at the W. M. Keck Observatory \citep{oke95,goo95,mcc98}, using two grisms of 300 $l$/mm and 400 $l$/mm.
The observations of 3C~323.1 (76 minutes in total) consisted of two sets of four wave-plate positions (0.0 deg, 22.5 deg, 45.0 deg, and 67.5 deg) $\times$ $7.5$ minute observations with 300 $l$/mm and a single set of four wave-plate positions $\times$ $4$ minute observations with 400 $l$/mm.
As noted in \cite{kis04}, the observed spectra at the wavelength range at $\lambda_{obs}\gtrsim$6500 \AA\ may be affected by second-order light contamination; therefore, the usable wavelength range for the analysis is $\lambda_{obs}=3500-6500$ \AA.
In this work we use the calibrated polarization spectra of 3C~323.1 presented in \cite{kis04}, which are kindly provided by M.~Kishimoto in electric form \citep[see][for details of the data analysis]{kis04}.
The ISP-corrected Keck/LRIS data used in this work are the same with those shown in Figure~9 of \cite{kis04}.
As listed in Table~6 of \cite{kis04}, the systemic polarization position angle $PA_{c}$ of 3C~323.1 at the epoch of the Keck/LRIS observation at the rest frame wavelength range of 4000-4731 \AA\ is $PA_{c} = 9.6$ deg.

Figure~\ref{fig:fig_pol} shows the ISP-corrected spectra of the rotated Stokes parameters $q'$ and $u'$, de-biased polarization degree $\hat{p}$, and $PA$ of 3C~323.1 obtained with Bok/SPOL \citep{sch00} and Keck/LRIS \citep{kis04}.
A discussion of the spectral variability of the polarization properties between these two measurements is given in Section~\ref{spec_var_pol_flux}.

To compare the Bok/SPOL and Keck/LRIS spectro-polarimetric measurements with the ESO3.6-m/EFOSC2 $V$-band imaging-polarimetry data, the $V$-band polarization degree $\hat{p}_V$ and the polarization position angle $PA_{V}$ are calculated from the Bok/SPOL and Keck/LRIS spectro-polarimetric data by convolving the total and polarized flux spectra ($F_{\lambda}$, $q_{\lambda}\times F_{\lambda}$ and $u_{\lambda}\times F_{\lambda}$) with the $V$-band filter transmission curve taken from \cite{bes90}.
The errors on $\hat{p}_V$ and $PA_{V}$ are evaluated as the sample standard deviation of the estimates from the 1000 trials of the Monte Carlo resampling of the spectra.
The $V$-band polarimetric properties of 3C~323.1 are summarized in Table~\ref{obs_Vband}.

The absolute flux calibration for the total flux (and the polarized flux) spectra of the two spectro-polarimetric measurements is discussed in Section~\ref{spectrophotometry}.

\subsubsection{Note on other historic polarimetric measurements for 3C~323.1}
\label{note_on_historic_pol}

Other than the three polarimetric data described above, we can find several historic polarimetric measurements in the literature \citep[][]{sto84,wil11}.
However, most of these measurements are obtained with white-light (no-filter) configuration, and thus, it is difficult to directly compare them with those obtained with Bok/SPOL, ESO3.6-m/EFOSC2, and Keck/LRIS because of the uncertainty of the wavelength coverage of the data; because the polarimetric properties of 3C~323.1 show wavelength dependence (see Section~\ref{spec_var_properties}), it is critical for the study of the polarimetric variability to use data taken in the same wavelength range.
Also, unlike the three polarimetric data used in this work, it is impossible to evaluate the magnitude values of the total flux of 3C~323.1 at the epochs of these white-light polarimetric data because of lack of referenceable photometric measurements.
For these reasons, we do not include the historic white-light polarimetric measurements for 3C 323.1 in the main text; they are instead summarized and discussed in Section~\ref{appendix_historic_whitelight}.

\subsection{Spectro-photometric flux calibration for the Bok/SPOL and Keck/LRIS data}
\label{spectrophotometry}

\begin{figure*}
\center{
\includegraphics[clip, width=6.4in]{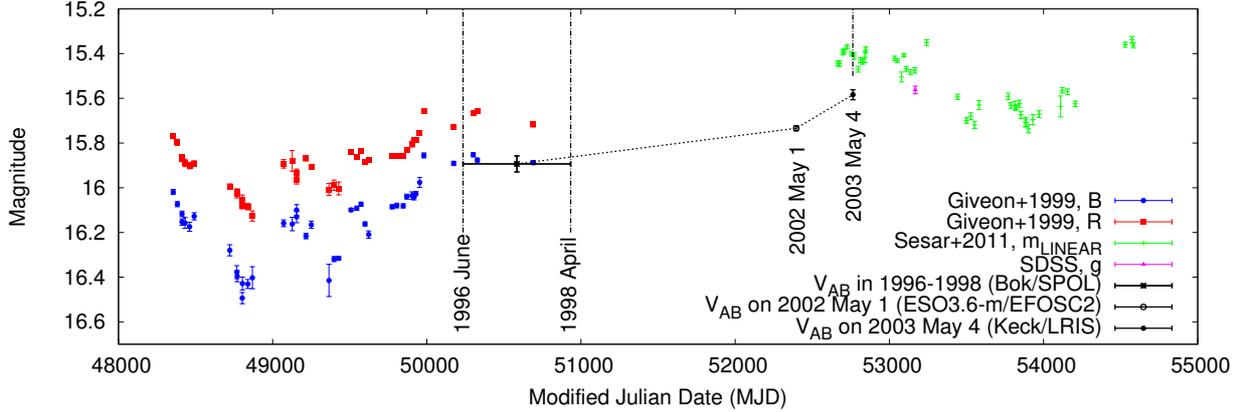}
}
 \caption{
 Compilation of the broad-band light curves for 3C~323.1.
 The $B$- and $R$-band light curves at MJD=48355-50690 are taken from Giveon et al. (1999), and the Lincoln Near-Earth Asteroid Research (LINEAR) survey recalibrated magnitudes ($m_{LINEAR}$) at MJD=52665-54582 are from  Sesar et al. (2011).
 The Sloan Digital Sky Survey $g$-band photometry at MJD=53168.32 is also plotted.
 The two periods of the Bok/SPOL spectro-polarimetry and the epoch of the Keck/LRIS spectro-polarimetry are indicated by dashed-dotted vertical bars.
 The estimated $V_{AB}$ magnitude at the epochs of the three polarimetric measurements (Bok/SPOL, ESO3.6-m/EFOSC2, and Keck/LRIS) is also shown (Table~\ref{obs_Vband}; see Section~\ref{eso_imapol} and Section~\ref{spectrophotometry} for details).
 }
 \label{fig:giveon_linear_lightcurve}
\end{figure*}

The absolute flux calibration of the spectroscopic data is generally not quite as accurate as that of the broad-band photometry data.
To confidently evaluate the variability of the total and polarized fluxes in 3C~323.1 at the epochs of the spectro-polarimetric observations by Bok/SPOL and Keck/LRIS described in Section~\ref{bol_keck_specpol}, we collect broad-band photometric data from the literature and publicly available databases (Figure~\ref{fig:giveon_linear_lightcurve}), and use them to estimate the $V$-band magnitude of the total flux.

\subsubsection{B- and R-band light curves in 1991-1998}
\label{wise_phot}

3C~323.1 was photometrically observed for seven years (1991-1998) in the Johnson-Cousins $B$- and $R$-band with the Wise Observatory 1-m telescope \citep[see][for details]{giv99}.
The $B$- and $R$-band light curves of 3C~323.1 presented by \cite{giv99}\footnote{\href{http://wise-obs.tau.ac.il/~givon/DATA/}{http://wise-obs.tau.ac.il/~givon/DATA/}} are shown in Figure~\ref{fig:giveon_linear_lightcurve}.
In the same figure, the epoch of \cite{sch00}'s Bok/SPOL spectro-polarimetric observations carried out in June 1996 and April 1998 is indicated by vertical bars.

Although there is no available broad-band photometric measurement obtained simultaneously with the Bok/SPOL data, we can infer the broad-band magnitude of 3C~323.1 at the periods of the Bok/SPOL observations from the photometry data of \cite{giv99}.
We assume that the weighted average of the four $R$-band measurements of \cite{giv99} from MJD=50174 to 50690, $R_{Vega}=15.6793\pm0.0353$ mag or $R_{AB}=15.7963\pm0.0358$ mag \citep[the AB offset and its uncertainty are taken from][]{fre94}, represents the magnitude value at the epoch of the observation of \cite{sch00}.
Here, we assume the uncertainty in $R_{Vega}$ as 0.0353 mag, which is derived not from the statistical error but from the sample standard deviation of the four measurements.

The Bok/SPOL total flux spectrum for 3C~323.1 is scaled so that the $R_{AB}$ magnitude calculated by convolving the Bok/SPOL total flux spectrum with the $R$-band filter transmission curve taken from \cite{bes90} coincides with $R_{AB}=15.7963$ mag.
Then, the polarized flux spectra are defined as the product of the polarization degree spectra in Figure~\ref{fig:fig_pol} and the scaled total flux spectrum.
The $V$-band magnitude of the scaled total flux Bok/SPOL spectrum is calculated as $V_{AB}=15.8935$ mag, by using the $V$-band filter transmission curve taken from \cite{bes90}.
Because the measurement errors on the Bok/SPOL total flux spectrum are negligible compared to the $R$-band magnitude uncertainty of 0.0358 mag, we adopt 0.0358 mag as the error of the estimated $V$-band magnitude.
The spectral monochromatic luminosity at 5100 \AA\ is $\lambda L_{\lambda} (5100$ \AA$)=1.1499 (\pm 0.0379)\times10^{45}$ erg/s.
The scaled Bok/SPOL total flux spectrum and the polarized flux spectrum are shown in Figure~\ref{fig:kecklris_flux}.

It should be noted that, since there is a time separation between the observations from the broad-band measurements in \cite{giv99} and the Bok/SPOL spectro-polarimetric measurements, the spectro-photometric calibration for the Bok/SPOL total and polarized flux spectra should have additional errors due to flux variability during the two Bok/SPOL observations in June 1996 and April 1998.
Nevertheless, the magnitude uncertainty of the Bok/SPOL data does not affect the final result of this work in the long term because the observational constraint on the polarization source from the well-correlated total flux and the polarized flux variability (discussed in Section~\ref{discuss}) can be verified even from only the two measurements of ESO3.6-m/EFOSC2 and Keck/LRIS whose magnitudes can be confidently evaluated.

\subsubsection{SDSS photometry and LINEAR light curve during 2003-2008}
\label{linear_phot}

3C~323.1 was photometrically observed by the SDSS Legacy Survey, using the SDSS imaging camera mounted on the Sloan Foundation 2.5-m telescope at the Apache Point Observatory \citep{gun98,yor00,gun06}.
We use the SDSS $g$-band PSF magnitude of 3C~323.1, $g=15.5606\pm0.0174$ mag, observed on the 12th June, 2004 (MJD=53168.3), retrieved from SDSS Data Release 12 (DR12) SkyServer \citep{ala15}.
We checked the SDSS photometry flag, and confirmed that the $g$-band photometry satisfies the ``clean photometry'' criteria recommended on the SDSS web page\footnote{\href{http://www.sdss.org/dr12/algorithms/photo_flags_recommend/}{http://www.sdss.org/dr12/algorithms/photo\_flags\_recommend/}}.

In addition, 3C~323.1 was observed between 2003 and 2008 as part of the Massachusetts Institute of Technology Lincoln Laboratory Lincoln Near-Earth Asteroid Research (LINEAR) survey using the 1-m LINEAR telescope \citep{sto00,ses11}.
The LINEAR observations were carried out with an unfiltered set-up, and the magnitude values stored in the LINEAR Survey Photometric Database \footnote{The LINEAR Survey Photometric Database is available at the SkyDOT Web site (\href{http://skydot.lanl.gov/}{http://skydot.lanl.gov/})} are given in the LINEAR recalibrated magnitude \citep{ses11}.

The SDSS photometry data and the LINEAR light curve for 3C~323.1 from MJD=52665 to 54582 are plotted in Figure~\ref{fig:giveon_linear_lightcurve}.
The multiple LINEAR measurements taken on the same night are binned to a weighted-average magnitude for each observation night.
In the same figure, the epoch of the Keck/LRIS spectro-polarimetric observation by \cite{kis04} on 4th May, 2003 (Section~\ref{bol_keck_specpol}), is indicated by a vertical bar.
As shown in Figure~\ref{fig:giveon_linear_lightcurve}, since the LINEAR photometry data were taken approximately simultaneously with both the SDSS data and the Keck/LRIS spectro-polarimetry data, we are able to use the LINEAR photometry data to estimate the broad-band magnitude of 3C~323.1 at the epoch of the Keck/LRIS observation, as described below.

Because quasar spectra are significantly different from stellar spectra due to their strong broad emission lines, we do not adopt the LINEAR-to-SDSS photometric transformation equation derived by using the SDSS stellar photometry in \cite{ses11}.
Instead, we assume that the LINEAR magnitude $m_{LINEAR}$ is related to the SDSS $g$-band magnitude as $g=m_{LINEAR}+m_{g,0}$.
This assumption is justified by the fact that the magnitude difference in the LINEAR measurements between the epochs of the observation of \cite{kis04} and the SDSS observation is as small as $\sim$0.07 mag, and thus the colour variability of 3C~323.1 between these epochs is also expected to be small \citep[see e.g.,][]{sch12}.
The LINEAR magnitude at MJD=53165.26 is $m_{LINEAR} = 15.4747 \pm 0.0123$ mag, which can be directly compared to the SDSS photometry obtained three days later, at MJD=53168.3, with the assumption that the quasar flux variability on a time-scale of several days is essentially negligible. 
From these values, the magnitude shifts are calculated as
\begin{eqnarray}
g&=&m_{LINEAR}+0.0859\ (\pm 0.0213)\rm{\ \ \ [mag]}.
\label{linear_trans}
\end{eqnarray}
Keck/LRIS spectro-polarimetry for 3C~323.1 was carried out at MJD=52763.6, and the LINEAR photometry data are available at MJD=52753.35 and MJD=52771.30 as $m_{LINEAR}=15.4014\pm0.0091$ mag and $m_{LINEAR}=15.4110\pm0.0140$ mag, respectively (the weighted-average is $m_{LINEAR}=15.4043\pm0.0076$ mag).
With the use of Equation~\ref{linear_trans}, the $g$-band magnitude of 3C~323.1 at the epoch of the Keck/LRIS observation can be estimated as 
$g=15.4902\pm 0.0226$ mag.

\begin{figure}
\center{
\includegraphics[clip, width=2.8in]{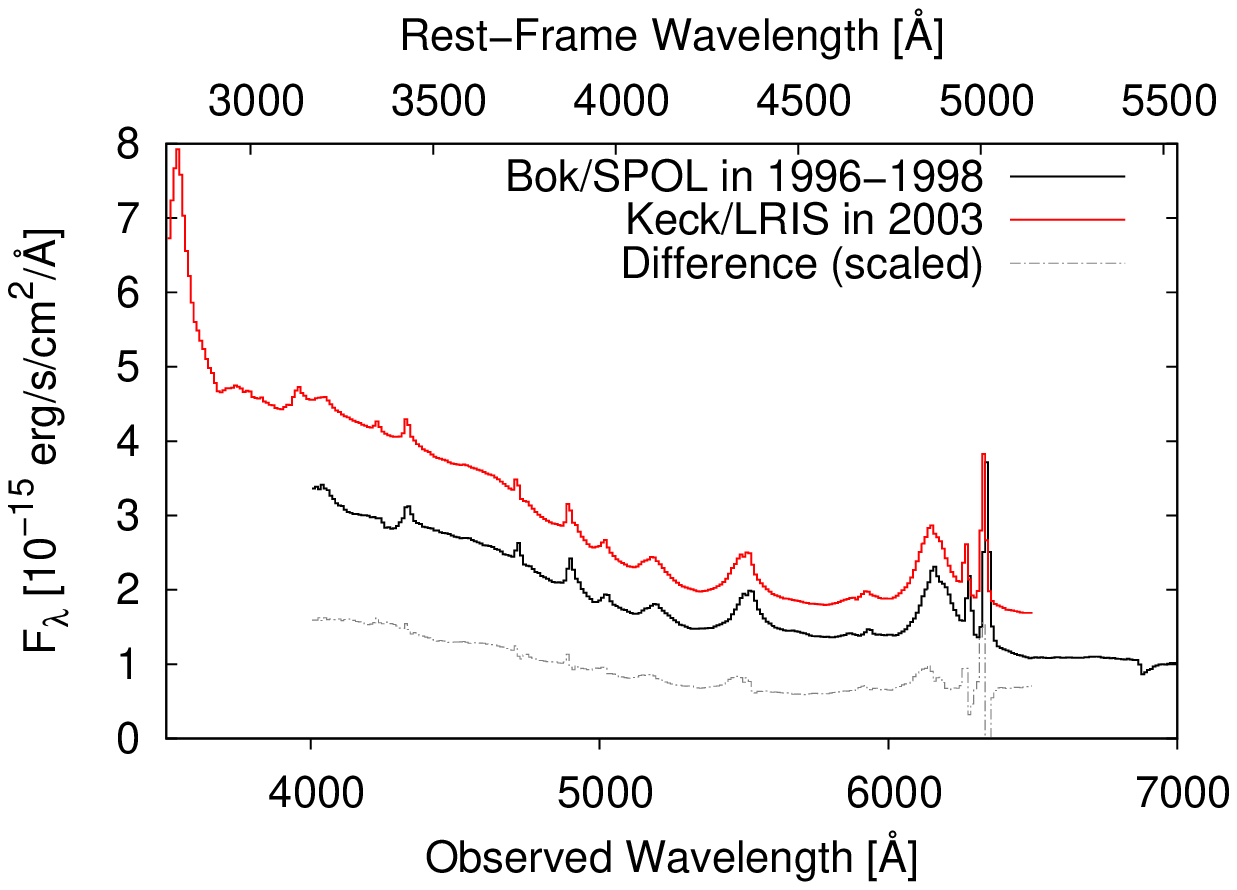}
\includegraphics[clip, width=2.8in]{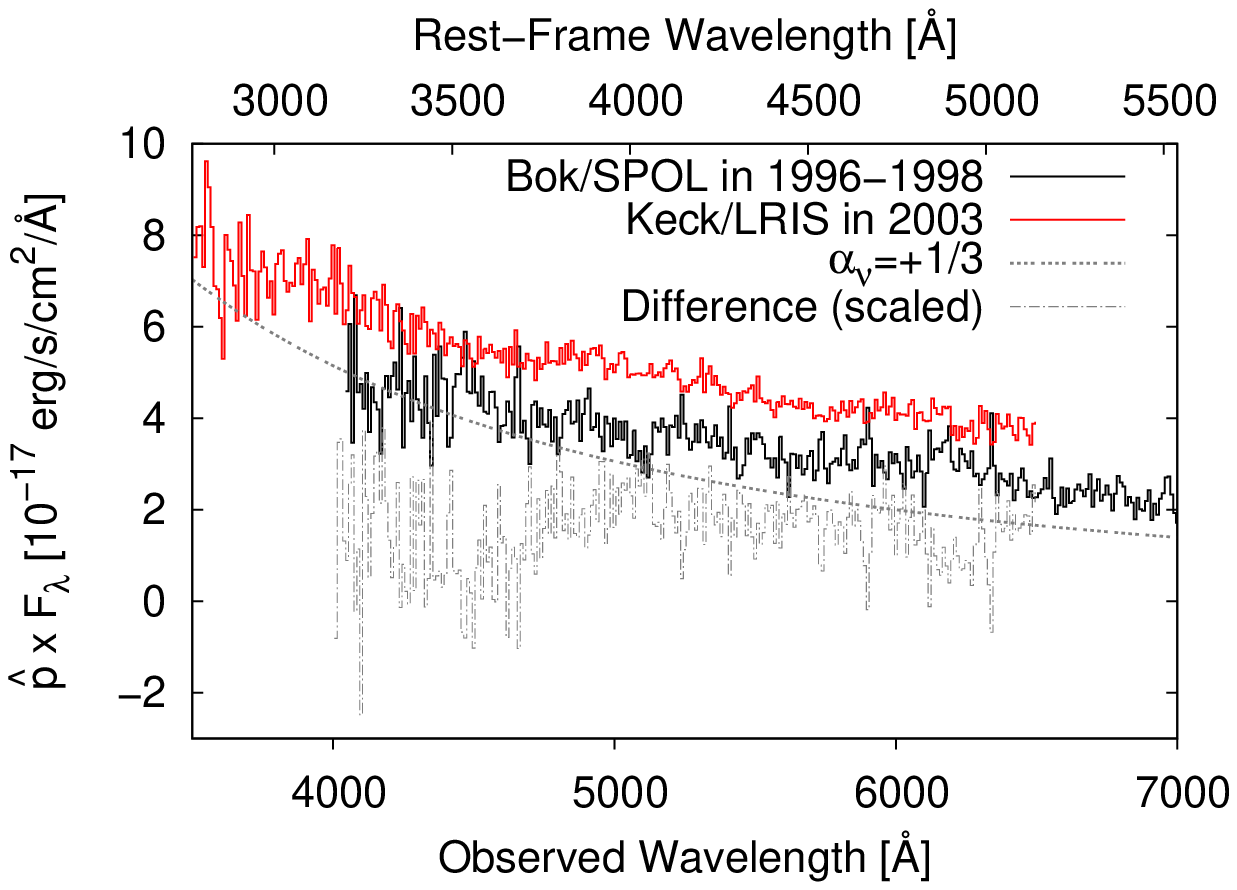}
}
 \caption{
 The total flux (top) and the polarized flux spectra (bottom) obtained with Bok/SPOL and Keck/LRIS.
 Galactic extinction is uncorrected.
 The arbitrarily-scaled difference spectra between the two measurements are also plotted.
 The spectra are binned into 10-\AA-wide bins. 
 For comparison, a power-law spectrum with $\alpha_{\nu}=+1/3$ is also plotted in the bottom panel.
 }
 \label{fig:kecklris_flux}
\end{figure}

The Keck/LRIS total flux spectrum for 3C~323.1 is scaled so that the $g$ magnitude calculated by convolving the Keck/LRIS total flux spectrum with the $g$-band filter transmission curve taken from \cite{doi10} coincides with $g=15.4902$ mag.
Then, the polarized flux spectra are defined as the product of the polarization degree spectra in Figure~\ref{fig:fig_pol} and the scaled total flux spectrum. 
The $V$-band magnitude of the scaled total flux Keck/LRIS spectrum is calculated as $V_{AB}=15.5838$ mag, by using the $V$-band filter transmission curve taken from \cite{bes90}.
Because the measurement errors on the Bok/SPOL total flux spectrum are negligible compared to the $g$-band magnitude uncertainty of 0.0226 mag, we adopt 0.0226 mag as the error of the estimated $V$-band magnitude.
The spectral monochromatic luminosity at 5100 \AA\ is $\lambda L_{\lambda} (5100$ \AA$)=1.7392(\pm 0.0362)\times10^{45}$ erg/s.
The scaled Keck/LRIS total flux spectrum and the polarized flux spectrum are shown in Figure~\ref{fig:kecklris_flux}.

\section{Spectral variability of the polarimetric properties}
\label{spec_var_properties}

In this section, the spectral variability between the two sets of spectro-polarimetry data obtained with Bok/SPOL and Keck/LRIS is examined.

\subsection{Spectral variability of the polarization degree and the polarization position angle}
\label{spec_var_p_pa}

Figure~\ref{fig:fig_pol} clearly shows that, as noted by \cite{sch00} and \cite{kis04}, the polarization degrees ($q'$, $u'$, and $\hat{p}$) at the wavelengths of the BLR emission are significantly diminished, making the polarization degree spectra strongly wavelength-dependent; in particular, the unpolarized ``small blue bump'', which is composed of the Balmer continuum and the UV Fe~II pseudo continuum from the BLR, is responsible for the decrease in the polarization degrees at $\lambda_{\rm{rest}}<4000$ \AA.
Although the spectra of polarization degree do not show strong variability between the two measurements, it should be noted that the polarization degree (i.e., $\hat{p}$) at the rest-frame wavelength range of $\lambda_{\rm{rest}}\sim3600$ \AA\ shows variations between the two spectro-polarimetric measurements (see Section~\ref{spec_var_pol_flux} for details).

Each $PA$ spectrum in the bottom panel of Figure~\ref{fig:fig_pol} is nearly wavelength-independent.
It is clear that, unlike the mainly time-constant polarization degrees, the $PA$ spectra show clear evidence of variability.
The difference in the $PA$ evaluated at the rest-frame wavelength range of 4000-4731 \AA\ is $\Delta PA_{c} = 4.6$ deg, and the $PA$ variability is nearly wavelength-independent (the bottom panel of Figure~\ref{fig:fig_pol}).
Because the systematic uncertainties in the grid of polarization standards are at most $\sim$ 1 deg \citep[e.g.,][]{sch92}, the difference in the observed $PA$ between the Bok/SPOL and Keck/LRIS measurements cannot be attributed to  errors of the $PA$ calibration by the use of different polarized standard stars, and therefore in this work the observed $PA$ variability is assumed to be the intrinsic variability of the polarization properties of 3C~323.1.

Although $PA$ variability is a general property of the blazar-like synchrotron emission \citep[e.g.,][]{ike11}, Section~\ref{add_note_synchrotron} shows that synchrotron emission is not responsible for the optical polarization observed in 3C~323.1.
If we assume that the optical polarization in 3C~323.1 is due to the scattering of the accretion disc continuum, the observed $PA$ variability requires some time-variable axi-asymmetric structure in the scattering region of 3C~323.1, as discussed in detail in Sections~\ref{const2} and ~\ref{const3}.

\subsection{Spectral variability of the polarized flux}
\label{spec_var_pol_flux}

In Figure~\ref{fig:kecklris_flux}, it is clear that the broad Balmer emission lines seen in the total flux spectra do not appear in either the Bok/SPOL or the Keck/LRIS polarized flux spectra; in other words, the BLR emission is unpolarized in 3C~323.1.
As discussed by \cite{kis04}, the Keck/LRIS polarized flux spectrum ($\hat{p}_{\lambda}\times F_{\lambda}$) shows a clear broad absorption-like feature at around $\lambda_{\rm{rest}}\sim3600$ \AA.
\cite{kis04} interpreted this feature as the Balmer continuum absorption, but they also noted the possibility that other higher-order Balmer series absorption lines and weak metal absorption lines also contribute to the absorption feature.
However, this spectral feature seems to be weak or absent in the Bok/SPOL $\hat{p}_{\lambda}\times F_{\lambda}$ spectrum.
Interestingly, the detailed shape of the two $\hat{p}_{\lambda}\times F_{\lambda}$ spectra also shows evidence of variability at the wavelength range of the H$\beta$ emission line ($\lambda_{\rm{rest}}\sim$4861 \AA); there is a broad absorption-like feature at around the wavelength range of the H$\beta$ emission line in the Keck/LRIS polarized flux spectrum, but it is not clearly seen in the Bok/SPOL polarized flux spectrum.
The spectral variability of $\hat{p}_{\lambda}\times F_{\lambda}$ at the wavelength ranges of the Balmer continuum and the H$\beta$ emission line can clearly be seen in the difference spectrum shown in the bottom panel of Figure~\ref{fig:kecklris_flux}.

The spectro-polarimetric variability of 3C~323.1 seen in the $\hat{p}_{\lambda}\times F_{\lambda}$ spectra (Figure~\ref{fig:kecklris_flux}) is very similar to that discovered by \cite{kis04} in Ton~202; \cite{kis04} have identified spectro-polarimetric variability between the two sets of Keck/LRIS data for Ton~202 obtained in 2002 and 2003 (one year apart), where the Balmer-edge absorption-like feature seen in the polarized flux spectrum in 2002 had disappeared by 2003.
\cite{kis04} suggested that the spectro-polarimetric variability of Ton~202 might be related to the time-variability of the geometry of the equatorial electron scattering region, but did not discuss how the changes in geometry resulted in the variability of the polarimetric properties around the Balmer continuum and emission lines.
Apparently, if the broad absorption features in the polarized flux spectrum are assumed to be intrinsic to the disc thermal emission, it is very difficult to explain reasonably the time-variability of the absorption features by the changes in the geometry of the scattering region alone.
Instead, as discussed in detail in Section~\ref{const3}, we propose that the absorption features are imprinted by an absorbing region with a time-variable structure, which is assumed to be spatially separated from the UV-optical emitting regions of the accretion disc.

In the next section (Section~\ref{discuss}), we focus on the $V$-band polarimetric and photometric variability.
As can be clearly seen in Figure~\ref{fig:kecklris_flux}, the polarized flux variability in the wavelength range of the $V$-band [$\lambda_{obs}\sim 5000-5900$ \AA\ at full width at half maximum (FWHM)] is clearly detected with a high signal-to-noise ratio.
It should be noted that the $V$-band is sampling the rest-frame wavelengths of $\lambda_{\rm{rest}}\sim 4000-4700$ \AA, within which the observed flux is dominated by the continuum emission.
Therefore, the $V$-band polarimetric variability mostly reflects the variability of the polarized continuum component.
This means that the final results of this work discussed in Section~\ref{discuss} are not affected by the putative changes to the polarization properties around the wavelength range of the Balmer continuum and other recombination lines discussed above.

\section{$V$-band polarimetric and photometric variability and its interpretation}
\label{discuss}

\begin{figure}
\center{
\includegraphics[clip, width=2.8in]{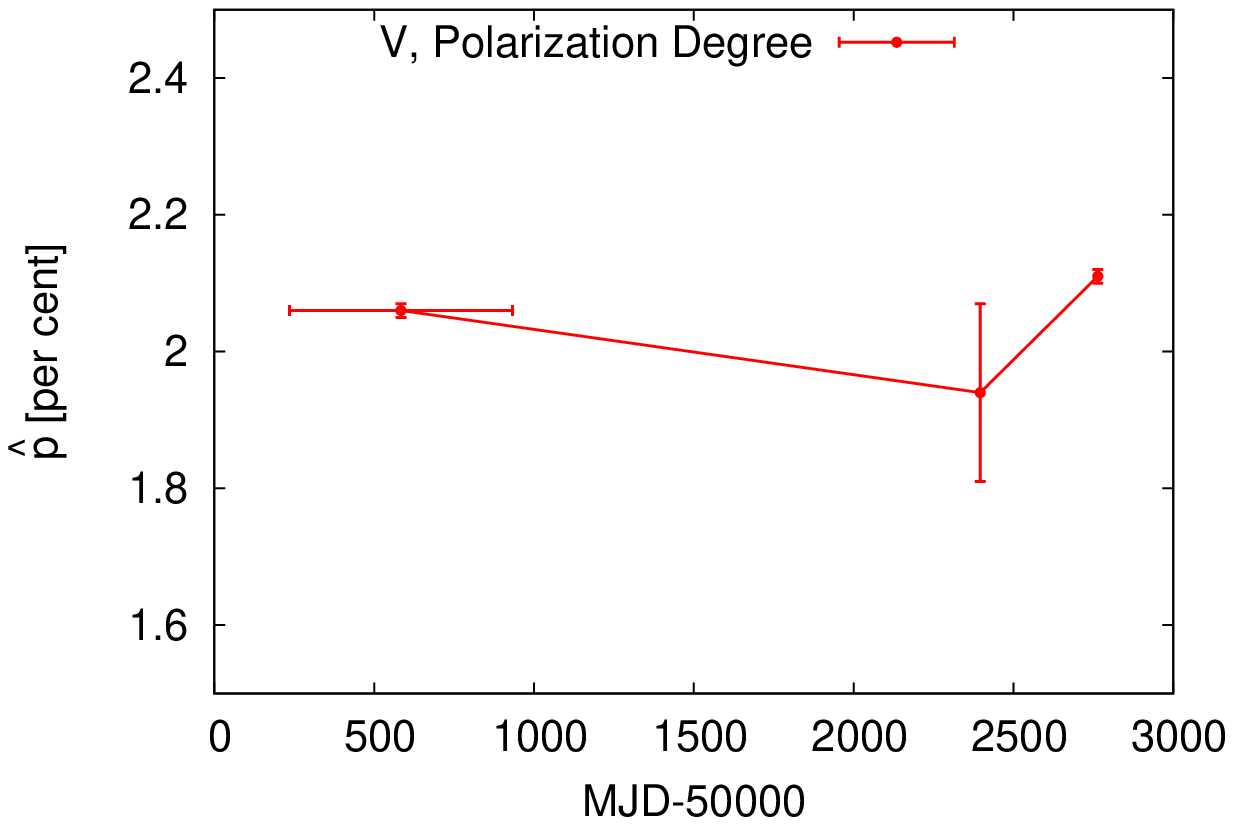}
\includegraphics[clip, width=2.8in]{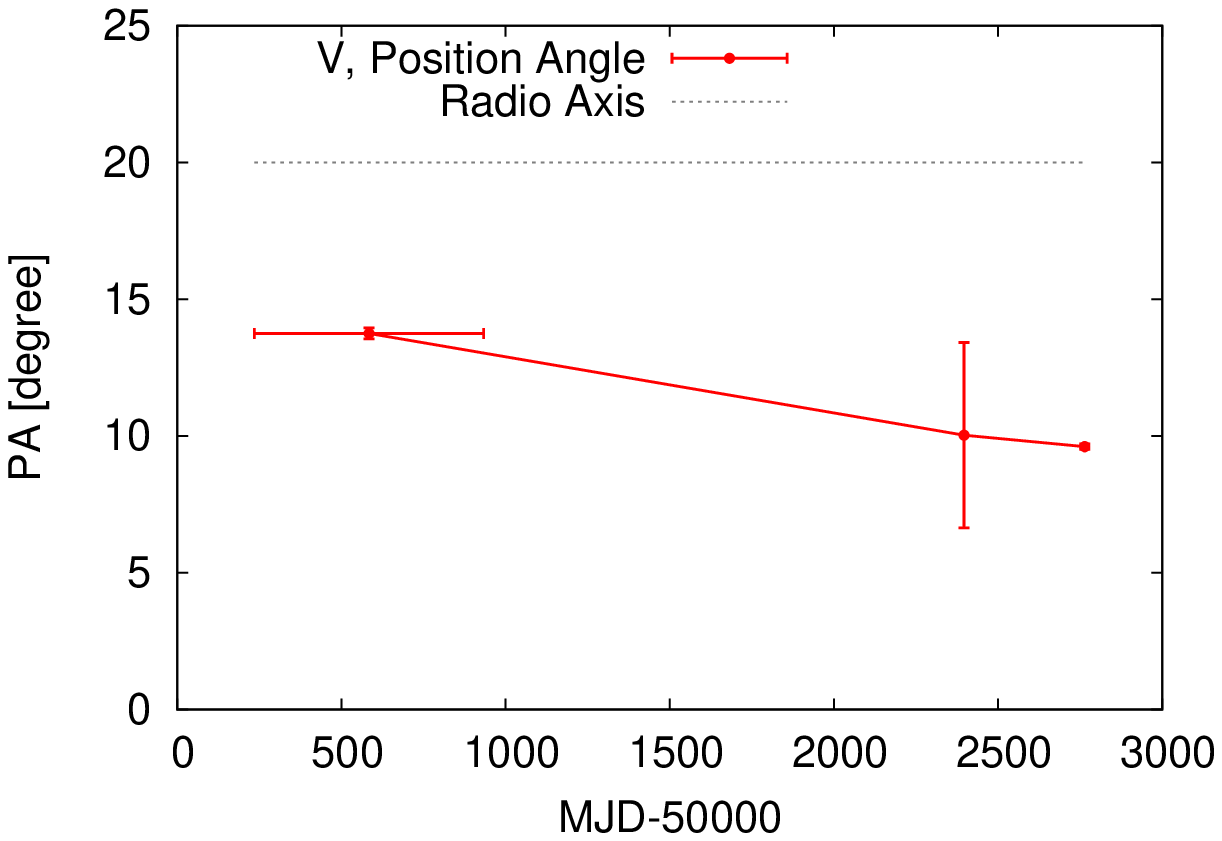}
\includegraphics[clip, width=2.8in]{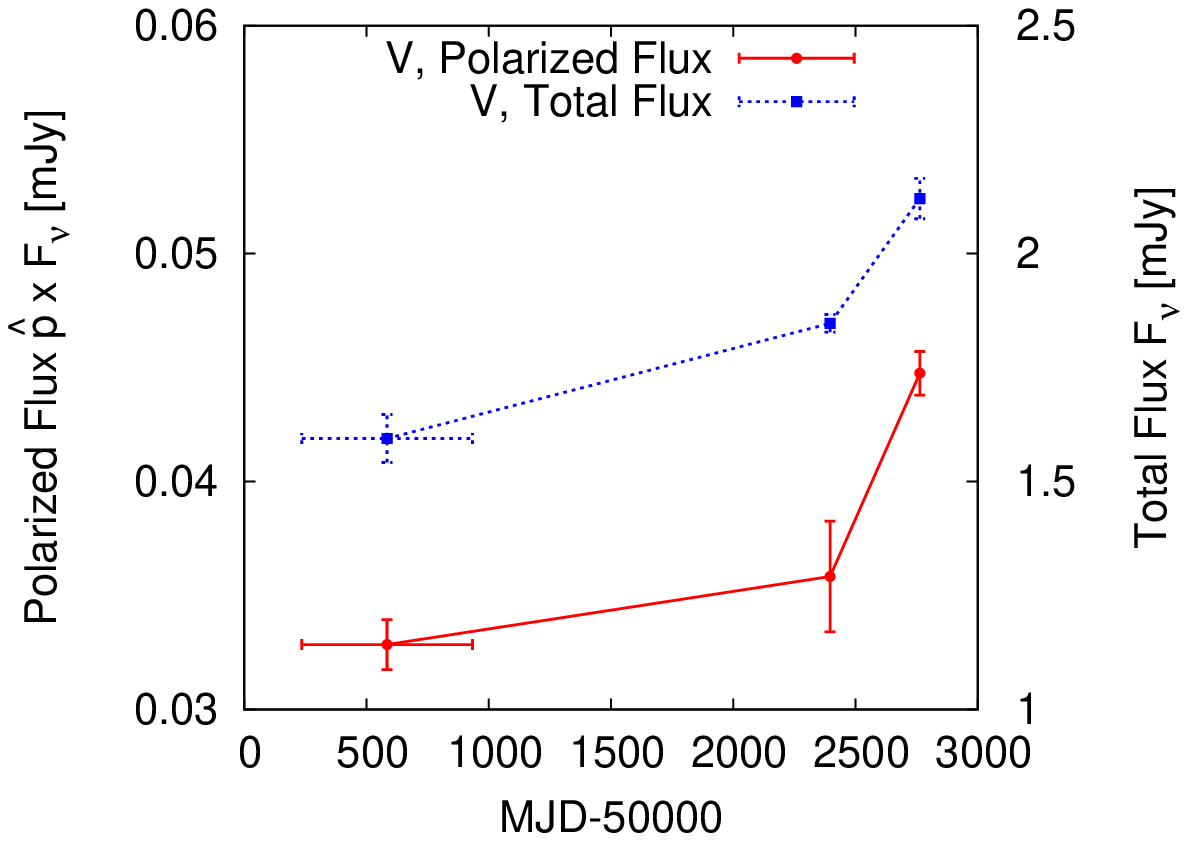}
}
 \caption{
 Top and middle: light curves of polarization degree $\hat{p}$ (top panel) and polarization position angle $PA$ (middle panel) of 3C~323.1 given in Table~\ref{obs_Vband}.
 3C~323.1's radio jet axis of $20$ deg \citep{kis04} is indicated as a dotted line in the middle panel.
 Bottom: the light curves of the $V$-band polarized flux $\hat{p}\times F_{\nu}$ (solid line) and total flux $F_{\nu}$ (dotted line) of 3C~323.1; Bok/SPOL in 1996-1998, ESO3.6-m/EFOSC2 on 1st May, 2002, and Keck/LRIS on 4th May, 2003.
 }
 \label{fig:summary_pol}
\end{figure}

Figure~\ref{fig:summary_pol} shows the photometric and polarimetric measurements for 3C~323.1 evaluated in the $V$-band summarised in Table~\ref{obs_Vband} as a function of time.
As shown in the top panel of Figure~\ref{fig:summary_pol}, the polarization degree $\hat{p}$ does not show strong variability during the three observations taken during the period 1996-2003.
The total flux $F_{\nu}$ of 3C 323.1 shows $\sim$ 0.3 mag variability during the same period, and the small variability in $\hat{p}$ results in the highly correlated variability between the $V$-band total flux and the polarized flux ($\hat{p}\times F_{\nu}$) (the bottom panel of Figure~\ref{fig:summary_pol}).

The middle panel of Figure~\ref{fig:summary_pol} shows the $V$-band $PA$, compared with 3C~323.1's radio jet axis of 20 deg \citep{kis04}.
As has already been noted in Section~\ref{spec_var_p_pa}, the observed $PA$ differs between the Bok/SPOL and Keck/LRIS measurements by $\sim 4$ deg, although the difference between the $PA$ and the radio axis is kept small (i.e., 3C~323.1 remains to be ``parallel'' polarization).

Below we discuss the geometrical constraints on the optical polarization source in 3C~323.1 derived from the polarimetric and photometric variability seen in Figure~\ref{fig:summary_pol}, and the possible interpretations of the optical polarization mechanism.

\subsection{Evidence against the synchrotron origin of the optical polarization in 3C~323.1}
\label{add_note_synchrotron}

As has already been mentioned in Section~\ref{spec_var_properties}, on the one hand, the Keck/LRIS polarized flux spectrum shows clear broad absorption features, but on the other hand, the Bok/SPOL polarized flux spectrum is almost a smooth power-law spectrum.
Based on the Bok/SPOL data, \cite{sch00} have suggested that the weak flux contribution of the synchrotron emission to the optical wavelength range can explain the observed optical polarization in 3C~323.1 (see Section~\ref{sec:intro} for details).
Even for the Keck/LRIS data, the synchrotron contribution scenario for the observed polarized flux cannot be ruled out with only the single epoch data; it is possible that the absorption features are imprinted in the intrinsically smooth synchrotron emission spectrum somewhere along the line of sight \citep{kis03}.

However, we point out here that the observed polarimetric and photometric variability strongly suggests that the optical synchrotron emission is not the origin of the polarized flux component in 3C~323.1.
As shown in Figure~\ref{fig:summary_pol}, the three $V$-band measurements reveal that both the total and the polarized fluxes vary almost simultaneously; in other words, the polarization degree is nearly time-constant (see also Figure~\ref{fig:fig_pol}).
Because the variable component of the total flux must be dominated by the disc continuum emission as discussed in Section~\ref{obs_properties}, the observed highly correlated variability between the total and polarized fluxes indicates that the polarized flux has a strong relationship with the accretion disc emission.
Therefore, the observation of the polarimetric and photometric variability essentially excludes the possibility that the optical polarization source in 3C~323.1 is the optical synchrotron emission.

In addition to the above mentioned evidence, with reference to the discussion provided in \cite{kis03} for the case of Ton~202, we can also confirm the difficulty in attributing the observed polarized flux in 3C~323.1 solely to the flux contribution from the optical synchrotron emission in terms of its core-only radio-loudness.
In general, the radio core emission of lobe-dominated quasars becomes optically thin at $\nu>30$ GHz \citep{ant90}, and, thus, the ratio of the radio core flux to the optical synchrotron flux (measured at $5000$ \AA) is expected to be Flux(radio core)/Flux(opt. sync.)$\sim$20,000 by assuming $f_{\nu}\propto \nu^{-1}$ \citep[][]{kis03}.
On the other hand, from 3C~323.1's radio-loudness for the core region $R_{\rm{core}}=41$ \citep{sch00} and the observed optical polarization degree $p\sim 2$\%, the flux ratio of the radio core flux to the optical polarized flux is $R_{\rm{core}}/p\sim41/0.02=$2050.
Therefore, if we assume that the optical polarized flux in 3C~323.1 originates from the synchrotron emission, the ratio of the radio core flux to the optical synchrotron flux must be Flux(radio core)/Flux(opt. sync.)$\sim$2050$\times p_{s}<$2050, where $p_{s}$ ($<$1) is the fractional polarization degree of the putative optical synchrotron radiation.
The discrepancy in the expected ($\sim$20,000) and required ($\sim$2050$\times p_{s}$) ratio of the radio core flux to the optical synchrotron flux implies that the amount of flux contribution from the optical synchrotron in 3C~323.1 is too small (by at least an order of magnitude) to account for the observed polarized flux.

Considering the lines of evidence discussed above, we conclude that the optical synchrotron emission is not relevant for the observed optical polarization in 3C~323.1.
In Section~\ref{const0}, we show that the observed polarimetric and photometric variability properties of 3C~323.1 are consistent with electron scattering from the equatorial scattering region, as proposed by \cite{kis04} (see Section~\ref{sec:intro}).
However, it is apparently difficult to account for the observed $PA$ variability seen in the middle panel of Figure~\ref{fig:summary_pol} by the simple equatorial scattering scenario.
We attempt to construct a geometrical model to explain all of the observed polarimetric and photometric properties of 3C~323.1 in Section~\ref{const0}.

\subsection{Constraints on the geometry of the scattering region in 3C~323.1}
\label{const0}

In this section, we first obtain observational constraints on the radial extent of the scattering region inferred from the polarimetric and photometric variability via causal reasoning, and compare them with the size of the BLR and the dust torus of 3C~323.1.
We then propose a geometrical model of the scattering/absorbing regions, which can potentially account for all of the observed polarimetric and photometric properties of 3C~323.1.

\subsubsection{Radial extent of the scattering region inferred from the polarimetric and photometric variability}
\label{const}

A robust constraint on the geometry of the scattering region can be derived from the highly correlated variability of the $V$-band total flux and the polarized flux.
As has already been pointed out in Section~\ref{add_note_synchrotron}, the two observations made by ESO3.6-m/EFOSC2 and Keck/LRIS on 1st May, 2002, and 4th May, 2003, respectively (separated by one year, i.e., by 0.8 years in the quasar rest frame), show coordinated variability between the total and polarized fluxes.
With the assumption that the polarization in 3C~323.1 is due to scattering of the disc continuum, the time lag between the total flux variation (disc continuum flux) and the polarized flux variation (scattered flux) corresponds to the light travel time across the accretion disc and the scattering region \citep[e.g.,][]{gas12}.
Therefore, the observed coordinated variability of the total and polarized fluxes directly constrains the radial distance of the scattering region as
\begin{eqnarray}
R_{\rm{sca}} (pF\ var.)&<&0.8 {\rm\ [light\ years]}.
\label{equation_pf}
\end{eqnarray}

On the other hand, within the framework of the equatorial scattering scenario, the $PA$ variability may possibly be interpreted as the change in geometry of the scattering region.
\cite{kis04} pointed out that the putative equatorial electron scattering region is assumed to be so small in size that its geometrical configuration can be changed within a one-year time-scale.
If the putative equatorial electron scattering region in 3C~323.1 is not an axisymmetric disc-like structure but has an axi-asymmetric clumpy density distribution, the $PA$ variability (with little variability in the polarization degree) may be a natural consequence of the orbital or bulk motion of the scattering region (see Section~\ref{const3} for details).
According to this interpretation, the $PA$ variability observed in 3C~323.1 during the Bok/SPOL and Keck/LRIS observations (the two observations are 5-7 years apart, i.e. a quasar rest frame time-lag of $<$ 6 years) implies that the maximum extent of the size of the scattering region $R_{\rm{sca}}$ is
\begin{eqnarray}
R_{\rm{sca}} (PA\ var.)&<&6 \times \left(\frac{v_{\rm{sca}}}{c}\right) {\rm\ [light\ years]}\nonumber\\
&=&0.20 \times \left(\frac{v_{\rm{sca}}}{10000 {\rm km/s}}\right) {\rm\ [light\ years]}
\label{equation_pa}
\end{eqnarray}
where $v_{\rm{sca}}$ represents the typical velocity of the polarization source.
If we assume that the polarization source is smaller in size than the BLR, it is reasonable to consider that the value of $v_{\rm{sca}}$ is larger than the velocity width of the broad emission lines.
Since the H$\beta$ broad emission line of 3C~323.1 has an FWHM of $\sim 7030$ km/s \citep[e.g.,][]{bor92}, we take $v_{\rm{sca}}=10,000$ km/s to be a reference value in Equation~\ref{equation_pa}.

\subsubsection{Comparisons of the radial extent between the polarization source, BLR, and the dust torus}
\label{const2}

The geometrical constraints on the scattering region $R_{\rm{sca}}$ derived in Section~\ref{const} should be compared with the radial extent of the accretion disc, BLR, and the dust torus.

Under the assumption of the \cite{sha73} accretion disc model, the disc radius at which the disc temperature matches the wavelength $\lambda_{\rm{rest}}$ as $k_B T_{\lambda_{\rm{rest}}}=hc/\lambda_{\rm{rest}}$ ($h$ and $k_B$ are the Planck constant and the Boltzmann constant, respectively) can be evaluated as \citep[e.g.,][]{mor10}
\begin{eqnarray}
R_{\rm{disc},\lambda_{\rm{rest}}} &\simeq& 0.01 \text{ [light years]}\nonumber\\
&\times& \left(\frac{\lambda_{\rm{rest}}}{\mu \text{m}} \right)^{4/3} \left(\frac{M_{\rm{BH}}}{10^9 M_{\odot}} \right)^{2/3} \left(\frac{L}{\eta L_E} \right)^{1/3},
\label{disc_model}
\end{eqnarray}
where $\eta\equiv L/(\dot{M}c^2)$ indicates the radiative efficiency of the disc.
The theoretical values of the radiative efficiency, including general relativistic corrections, range from 6\% to 42\% as a monotonically increasing function of the black hole spin \citep[e.g.,][]{sha83,fra92,kat08}.
By substituting $\log (M_{\rm{BH}}/M_{\odot})=9.07$ and $L/L_E=0.10$ taken from \cite{she11} (see Section~\ref{obs_properties}), the disc radius of $\lambda_{\rm{rest}}=5100$ \AA\ in 3C~323.1 can be calculated to be within the range
\begin{equation}
R_{\rm{disc},5100\text{\AA}}= 0.003 - 0.005 {\rm\ [light\ years]}.
\label{disc_radius}
\end{equation}
Recent observations of quasar microlensing events and AGN continuum reverberation mapping \citep[e.g.,][]{mor10,ede15} have revealed that the actual accretion disc sizes in quasars are larger by a factor of $\sim$4 than the size predicted by the \cite{sha73} accretion disc model, but even considering such uncertainty, we can conclude from Equation~\ref{disc_radius} that the accretion disc radii responsible for the optical emission in 3C~323.1 can be regarded as a point source when viewed from the BLR or the dust torus (see below).

The H$\beta$ BLR radius has been determined by reverberation mapping of the broad H$\beta$ emission lines for several tens of AGN/quasars, and can be well expressed as a function of the continuum luminosity as \citep{ben09}
\begin{equation}
\log R_{\text{BLR}} [\text{light days}] = -21.3 + 0.519 \log(\lambda L_{\lambda}(5100 \text{\AA})\text{ [erg/s]}) 
\end{equation}
By substituting the monochromatic luminosity $\lambda L_{\lambda}(5100 \text{ \AA})$ at the epochs of Bok/SPOL and Keck/LRIS (see Section~\ref{spectrophotometry}), the BLR radius of 3C~323.1 is found to be in the range of
\begin{equation}
R_{\text{BLR}} = 0.33 - 0.41 {\rm\ [light\ years]}.
\label{blr_size}
\end{equation}

The dust reverberation radius of AGN/quasars \citep[e.g.,][and references therein]{bar92,sug06,kis11,kos14}, which represents the radius of the innermost dust torus, is known to be well represented as a function of the continuum luminosity as $\log R_{\text{dust, in}} \text{ [pc]} = -0.88 + 0.5 \log(\lambda L_{\lambda}(5500 \text{ \AA})/10^{44}\text{ [erg/s]})$ \citep{kos14}.
By assuming $f_{\nu}\propto \nu^{0}$, 
this relation can be converted to 
\begin{equation}
\log R_{\text{dust, in}} [\text{light days}] = -19.8 + 0.5 \log(\lambda L_{\lambda}(5100 \text{\AA})\text{ [erg/s]})
\end{equation}
By substituting the monochromatic luminosity at the epochs of Bok/SPOL and Keck/LRIS, the innermost radius of 3C~323.1's dust torus is in the range of
\begin{equation}
R_{\text{dust, in}} = 1.47 - 1.81 {\rm\ [light\ years]}.
\end{equation}

\begin{figure}
\center{
\includegraphics[clip, width=2.8in]{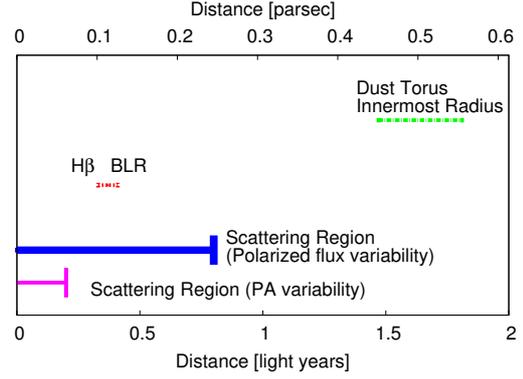}
}
 \caption{Observational constraints on the radial extent of the scattering region in 3C~323.1.
 Two independent upper limits on the radial extent of the scattering region are taken from the coordinated variability of the total flux and the polarized flux (Equation~\ref{equation_pf}) and the observed time-scale of the polarization position angle variability (Equation~\ref{equation_pa} assuming $v_{\rm{sca}}=10,000$ km/s).
 The possible location of the H$\beta$ broad line region (BLR) and the radius of the innermost dust torus are also shown (see Section~\ref{const2} for details).
Note that the optical emitting region of the accretion disc in 3C~323.1 can be regarded as a central point source if viewed from the BLR and the dust torus (see Equation~\ref{disc_radius}).}
 \label{fig:distance}
\end{figure}

Figure~\ref{fig:distance} shows the comparison of the estimated radius of the BLR and the dust torus innermost radius of 3C~323.1 with the upper limits of the radial extent of the scattering region given in Equations~\ref{equation_pf} and \ref{equation_pa}.
From this comparison, we can firmly conclude that the scattering region is located inside the dust torus.
This means that scattering from regions larger than the dust torus as the dominant polarization mechanism in 3C~323.1 can be completely ruled out.
Because the innermost radius of the dust torus corresponds to the sublimation radius of the dust grains \citep[e.g.,][]{kis07,mor12,kos14,net15}, there must be no dust grains inside the dust torus, and thus we can also conclude that electron scattering is the main contributor to the observed optical polarization in 3C~323.1.
Moreover, by adopting the size constraint of $R_{\rm{sca}} (PA\ var.)<0.20$ light years (Equation~\ref{equation_pa}) at face value, the scattering region should be smaller in size than the BLR, which in turn validates the assumption of the high velocity of the scattering region $v_{\rm{sca}}$ adopted to derive the constraint on $R_{\rm{sca}}$ in Equation~\ref{equation_pa}.
As has been mentioned in Section~\ref{sec:intro}, the scattering region inside the BLR cannot produce the net polarization of the BLR emission, ensuring the null-polarization of the BLR emission of 3C~323.1.
In summary, the constraints on the radial extent of the scattering region in 3C~323.1 obtained in Section~\ref{const} are consistent with the equatorial electron scattering scenario proposed by \cite{kis03,kis04,kis08} as the interpretation of the polarization properties of quasars with continuum-confined polarization.

\subsubsection{A possible model for the scattering geometry in 3C~323.1}
\label{const3}

Although the constraints on the radial extent of the scattering region in 3C~323.1 obtained in Section~\ref{const2} suggest that the scattering region is located inside the BLR, they do not specify the detailed geometry of the scattering region. 
Indeed, as was partially discussed above (Sections~\ref{spec_var_properties} and \ref{const}), several lines of evidence suggest that the simple assumption that there is an axisymmetric disc-like electron-scattering region that directly scatters accretion disc continuum photons into our line of sight cannot fully account for the observed polarimetric and photometric properties of 3C~323.1.

To infer the true geometrical structure of the inner region of 3C~323.1, we have to take the following observational properties into account:
\begin{itemize}
\item As noted in Section~\ref{const}, the observed $PA$ variability in 3C~323.1 requires the presence of some axi-asymmetric structure in the scattering region, which changes its geometry on a time-scale of several years;
\item The variability of the broad absorption like features in the polarized flux spectra between the Bok/SPOL and the Keck/LRIS observations (Section~\ref{spec_var_pol_flux}) suggests that either (a) the broad absorption features are intrinsic to the accretion disc emission and the strength of the absorption features are time-variable, or (b) there is a high-velocity absorbing region somewhere along the line of sight whose geometry is time-variable;
\item As noted in Section~\ref{sec:intro} (see also Section~\ref{obs_properties}), \cite{kok16} has shown that the spectral shape of the intrinsic accretion disc continuum of 3C~323.1 revealed as the variable component spectrum is inconsistent with that revealed as the polarized flux spectrum, in that the variable component spectrum is much bluer than the polarized flux spectrum (Figure~\ref{fig:kecklris_flux}).
Within the equatorial scattering geometry, the discrepancy of the spectral shape between the variable and the polarized flux components can only be explained by the above-mentioned scenario (b) as long as the absorbing materials are co-spatial with the equatorial scattering region or located between the accretion disc and the scattering region.
\end{itemize}

\begin{figure*}
\center{
\includegraphics[clip, width=5.4in]{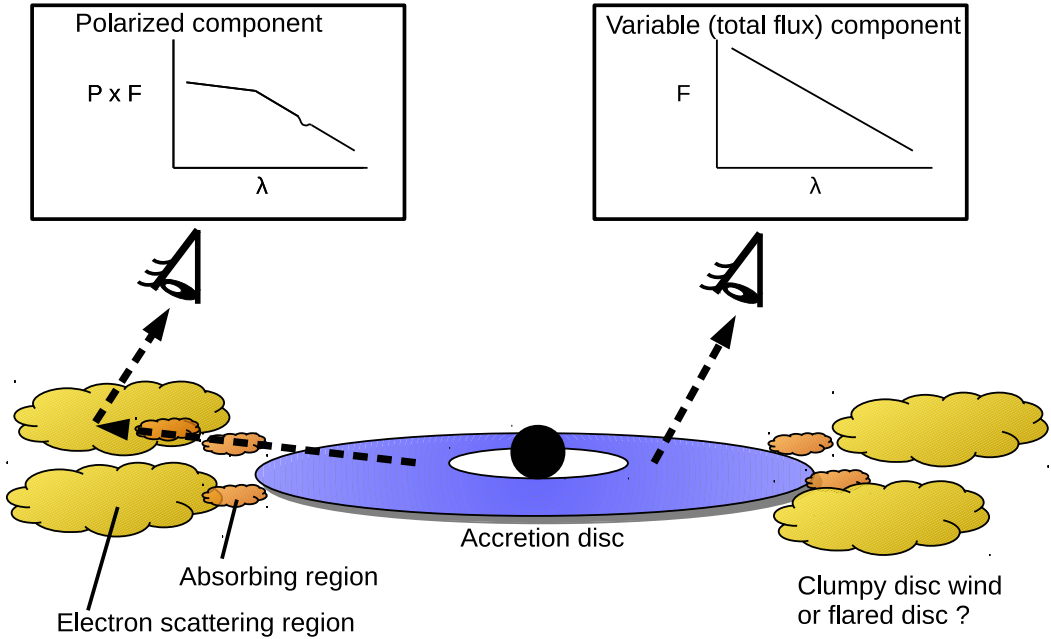}
}
 \caption{
 Schematic drawing of the geometrical model for the scattering region of 3C~323.1 and of other quasars with continuum-confined polarization.
 Note that the broad line region (BLR) is situated outside the optically-thin equatorial electron-scattering region, ensuring the null-polarization of the BLR emission.
 The variable component spectrum (i.e., the total flux spectrum) directly reflects the spectral shape of the disc continuum emission, while the polarized flux spectrum shows broad absorption features (the Balmer continuum and possibly the Balmer and metal lines; see Figure~\ref{fig:kecklris_flux}) due to the equatorial absorbing region, which is either co-spatial with the scattering region or located between the accretion disc and the scattering region.
 The geometric configurations of both the scattering and the absorbing regions are assumed to be axi-asymmetric and time-variable to account for the slight variations of the polarization position angle and the absorption features in the polarized flux observed in 3C~323.1.
A clumpy disc wind or flared disc surface may possibly be the physical origin of the absorbing/scattering regions.
 }
 \label{fig:model}
\end{figure*}

Figure~\ref{fig:model} illustrates the geometrical model that can explain the above-mentioned observational properties of 3C~323.1.
In this model, an equatorial absorbing region responsible for the broad absorption features (the Balmer continuum, and possibly the Balmer and other weak metal lines, as mentioned in Section~\ref{spec_var_pol_flux}) is located inside (or co-spatial with) the optically-thin equatorial electron scattering region.
With this geometrical configuration, the total flux spectrum, and similarly the variable component spectrum, directly reflects the spectral shape of the intrinsic accretion disc continuum, while the polarized flux spectrum shows additional broad absorption features induced by the equatorial absorbing region because the observed polarized light has once passed through the absorbing region.
This also indicates that the intrinsic accretion disc continuum is featureless as observed as the variable component spectrum.
The geometrical model in Figure~\ref{fig:model} is essentially the same as that considered by \cite{kis03} and \cite{kok16} [see the model (a) in Figure~6 of \citealt{kis03}], but in our model the structure of the absorbing/scattering regions is assumed to be axi-asymmetric and time-variable to account for the polarimetric variability observed in 3C~323.1.
By requiring that the radial extent of the scattering region is smaller in size than the BLR (i.e., $R_{\rm{sca}}\lesssim 0.33-0.41$ light years; Equation~\ref{blr_size}), the dynamical time-scale (Kepler orbital time) $t_{dyn}$ of the scattering materials is at most $t_{dyn} \lesssim 88-122 {\rm\ years}$, assuming $\log (M_{\rm{BH}}/M_{\odot})=9.07$.
Therefore, it is possible, in principle, that some dynamical motions (not only the orbital motions but also the radial motions) of the scattering materials gradually modify the net polarization $PA$ on time-scales of several years while keeping the polarization degree nearly constant, as observed in 3C 323.1. 
the spectral variability of the polarized flux spectra between the Bok/SPOL and Keck/LRIS observations discussed in Section~\ref{spec_var_pol_flux} can be attributed to some dynamical motions of the absorbing materials.

From the currently available observational constraints alone, it is impossible to specify the physical origins of the absorbing/scattering materials assumed in the proposed model.
One promising candidate to produce time-variable inhomogeneous equatorial scattering/absorbing structures is the accretion disc wind, similar to those observed in broad absorption line (BAL) quasars \citep{elv00,pro04,yon07,mat16}, especially Balmer-BAL quasars \citep[][and references therein]{zha15}.
In addition, it may be possible that the accretion disc itself is flaring up at a particular radius \citep[e.g.,][]{lir11,jia16} and that the flared region absorbs some quantity of photons from the inner disc region travelling toward the equatorial direction.

Although this work deals with only a single radio-loud quasar 3C~323.1, we believe that the optical polarization source in other radio-quiet/radio-loud quasars with continuum-confined polarization must be similar to that proposed here considering the similarities of the spectral shape of the polarized flux spectra and the variable component spectra \citep[see][]{kis04,kis08,kok16}.
Further polarimetric and multi-wavelength monitoring observations for quasars with continuum-confined polarization will eventually clarify the detailed geometry and the true nature of the scattering/absorbing regions in these quasars.

%
%
%
%

\section{Summary and conclusions}
\label{summary_and_conclusion_3c3231}

We examine the polarimetric and photometric variability of 3C~323.1 by using two optical spectro-polarimetric measurements taken during 1996-1998 (Bok/SPOL) and 2003 (Keck/LRIS) and a $V$-band imaging-polarimetric measurement taken in 2002 (ESO3.6-m/EFOSC2).
The two spectro-polarimetric measurements reveal that the polarimetric properties at the wavelength range of the Balmer continuum and other recombination lines are time-variable, while the polarized degree $\hat{p}$ of the continuum emission is nearly constant (Figures~\ref{fig:fig_pol} and \ref{fig:kecklris_flux}).
The $V$-band polarimetric and photometric measurements (where two of the three measurements are calculated from the two spectro-polarimetry data sets) confirm that the polarization $PA$ is variable on a time-scale of several years (Figure~\ref{fig:summary_pol}).
Moreover, the light curves of the $V$-band total and polarized fluxes show highly correlated variability (Figure~\ref{fig:summary_pol}), indicating that the polarized flux is the scattered disc continuum emission because there is evidence to suggest that the variability of the total flux is due to the intrinsic variability of the accretion disc continuum emission.

The variability time-scales of the $PA$ and the polarized flux introduce the constraint that the polarization source in 3C~323.1 is located inside the BLR (Figure~\ref{fig:distance}), which is consistent with the equatorial electron scattering scenario suggested by \cite{kis04}.
We propose a more sophisticated geometrical model to account for the polarimetric and photometric properties, as well as the variability in behaviour, observed in 3C~323.1.
Our model comprises an equatorial absorbing region and an optically-thin equatorial electron-scattering region, which surrounds the UV-optical emitting regions of the accretion disc (Figure~\ref{fig:model}).
The absorbing region is assumed to be co-spatial or smaller in size compared to the scattering region.
As schematically shown in Figure~\ref{fig:model}, this geometrical configuration can explain the discrepancy in the spectral shape of the accretion disc continuum emission in 3C~323.1 revealed as the variable component spectrum \citep[measured by][]{kok16} and as the polarized flux spectrum.
The structures of both the absorbing region and the scattering region are assumed to be axi-asymmetric and time-variable to account for the variability of the broad absorption features in the polarized flux and the variability of the $PA$ observed in 3C 323.1.

Although the physical origins of the absorbing region and the scattering region are unclear, the accretion disc wind must be a promising candidate to produce the equatorial absorbing/scattering regions around the accretion disc.
We believe that the structure of the inner regions of other quasars with continuum-confined polarization must be similar to that proposed here for 3C~323.1, considering the similarities in the spectral shape of the polarized flux spectra and the variable component spectra \citep[see][]{kis04,kis08,kok16}.
Further polarimetric and multi-wavelength monitoring observations for these quasars will enable us to probe the true nature of the internal structure of these quasars.

\section*{Acknowledgements}

This work was supported by JSPS KAKENHI Grant Number 15J10324.
We thank Gary D. Schmidt, Paul S. Smith, and Makoto Kishimoto for providing us with the reduced spectro-polarimetric data presented in \cite{sch00} and \cite{kis04}.
This research has made use of NASA's Astrophysics Data System Bibliographic Services.
This work is partially based on data obtained from the ESO Science Archive Facility (ESO Programme ID 68.A-0373, PI: D. Hutsem\'ekers).

The LINEAR program is sponsored by the National Aeronautics and Space Administration (NRA No. NNH09ZDA001N, 09-NEOO09-0010) and the United States Air Force under Air Force Contract FA8721-05-C-0002.

Funding for SDSS-III has been provided by the Alfred P. Sloan Foundation, the Participating Institutions, the National Science Foundation, and the U.S. Department of Energy Office of Science. The SDSS-III web site is http://www.sdss3.org/.

SDSS-III is managed by the Astrophysical Research Consortium for the Participating Institutions of the SDSS-III Collaboration including the University of Arizona, the Brazilian Participation Group, Brookhaven National Laboratory, Carnegie Mellon University, University of Florida, the French Participation Group, the German Participation Group, Harvard University, the Instituto de Astrofisica de Canarias, the Michigan State/Notre Dame/JINA Participation Group, Johns Hopkins University, Lawrence Berkeley National Laboratory, Max Planck Institute for Astrophysics, Max Planck Institute for Extraterrestrial Physics, New Mexico State University, New York University, Ohio State University, Pennsylvania State University, University of Portsmouth, Princeton University, the Spanish Participation Group, University of Tokyo, University of Utah, Vanderbilt University, University of Virginia, University of Washington, and Yale University.

%
%

\bibliography{./qpv.bib}


\appendix  

\section{Correcting for the Effects of the Galactic ISP}
\label{appendix_isp_correction}

\begin{figure}
\center{
\includegraphics[clip, width=3.2in]{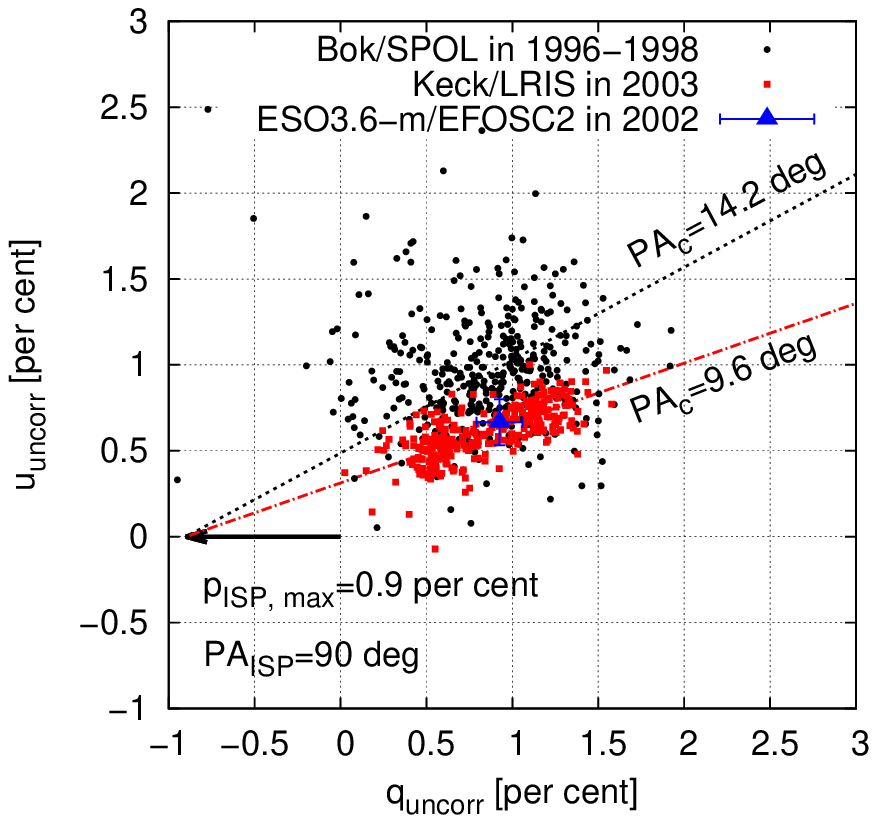}
}
 \caption{The ISP-uncorrected $q-u$ diagram of the Bok/SPOL and Keck/LRIS spectro-polarimetry data, and the ESO3.6-m/EFOSC2 $V$-band polarimetry data for 3C~323.1.
 For the spectro-polarimetry data, each data point is the $q-u$ value at a wavelength bin of 10 \AA\ width.
 The thick arrow indicates the direction of the ISP polarization, and the dotted lines indicate the ISP-corrected systemic polarization position angles $PA_c$ of the two spectro-polarimetry data evaluated at the rest-frame wavelength range of 4000-4731\AA.}
 \label{fig:qu_2dplot}
\end{figure}

\cite{kis04} pointed out that the observed optical polarization of 3C~323.1 is affected by a non-negligible amount of the Galactic ISP.
The wavelength dependence of the ISP linear polarization can be described by the empirical formula \citep[e.g.,][]{ser75}
\begin{equation}
p_{\text{ISP}}=p_{\text{ISP, \text{max}}}\exp\left[ -K \ln^2\left(\frac{\lambda_{\text{max}}}{\lambda}\right) \right],
\label{equation_isp}
\end{equation}
where $K=0.01+1.66\lambda_{\text{max}}$ \citep{whi92}.
In this work, we adopt the ISP parameters of $p_{\text{ISP, max}}=0.90$\% and $\lambda_{\text{max}}=5570$\AA, and the ISP position angle of $PA_{\text{ISP}}=90$ deg derived by \cite{kis04} for 3C~323.1.
$PA_{\text{ISP}}=90$ deg means that the ISP vector in the direction of 3C~323.1 is parallel to the $q$-axis in the $q-u$ plane, and thus, the ISP correction for $q$ and $u$ can be described as $q=q_{\text{uncorr}}+p_{\text{ISP}}$ and $u=u_{\text{uncorr}}$, where $q_{\text{uncorr}}$ and $u_{\text{uncorr}}$ represent the ISP-uncorrected values (see Figure~\ref{fig:qu_2dplot}).

Figure~\ref{fig:qu_2dplot} shows the ISP-uncorrected $q-u$ (i.e., $q_{\text{uncorr}}-u_{\text{uncorr}}$) diagram of the Bok/SPOL and Keck/LRIS spectro-polarimetry data, and the ESO3.6-m/EFOSC2 $V$-band imaging-polarimetry data for 3C~323.1.
In Figure~\ref{fig:qu_2dplot}, the ISP vector of $p_{\text{ISP}}=0.9$\% is shown as a representative value, but it should be noted that the actual ISP correction is slightly wavelength-dependent (Equation~\ref{equation_isp}).
The ISP correction leads to a shift of the origin of the intrinsic polarization vector to $(q_{\text{uncorr}}, u_{\text{uncorr}})=(-p_{\text{ISP}},0)$, and the ISP-corrected $PA$ is defined as half the geometrical angle between the $q$-axis and lines passing through the shifted origin and data points.
As indicated in Figure~\ref{fig:qu_2dplot}, the Keck/LRIS data points are linearly aligned along the vector of $PA_c=9.6$ deg, meaning that the ISP-corrected $PA$ of 3C~323.1 is nearly wavelength-independent \citep[Figure~\ref{fig:fig_pol};][]{kis04}.
As we can see in Figure~\ref{fig:fig_pol}, the ISP-corrected $PA$ spectrum obtained with Bok/SPOL is also nearly wavelength-independent.

\section{Historic white-light polarimetric measurements for 3C~323.1}
\label{appendix_historic_whitelight}

Historic polarimetric measurements for 3C~323.1 presented in \cite{sto84} and \cite{wil11} are summarized in Table~\ref{obs_unfilter}.
It should be noted that, unlike the Bok/SPOL, ESO3.6-m/EFOSC2, and Keck/LRIS polarimetric measurements given in Table~\ref{obs_Vband}, it is impossible to estimate the $V$-band magnitudes of the total flux at each epoch of the polarimetric measurement summarized in Table~\ref{obs_unfilter} because referenceable photometric measurements at these epochs are not available in the literature.
Also, we do not correct for the Galactic ISP for these historic data, because it is difficult to evaluate the ISP strength from Equation~\ref{equation_isp} due to the uncertainty of the wavelength coverage (see below).

In Figure~\ref{fig:summary_pol_nonfilter}, the time variation of the ISP-uncorrected $\hat{p}$ and $PA$ of 3C~323.1 as given in Table~\ref{obs_unfilter} are plotted, where different symbols indicate different filter configurations.
We also plot the ISP-uncorrected $\hat{p}_{V}$ and $PA_{V}$ derived from the Bok/SPOL, ESO3.6-m/EFOSC2, and Keck/LRIS data in the same figure.
As has already been pointed out in Section~\ref{note_on_historic_pol}, the historic polarimetric measurements summarized in Table~\ref{obs_unfilter} are not taken with standard photometric filter system.
Considering the uncertainty of the wavelength coverage of each measurement, what we can only infer about the variability of $\hat{p}$ in 3C~323.1 from the historic (mostly white-light) polarimetric data in Table~\ref{obs_unfilter} is that there is no evidence to support the existence of strong variability of $\hat{p}$ for the last several decades.
On the other hand, since the wavelength-dependence of $PA$ of 3C~323.1 is relatively weak, it may be valid to conclude that the $PA$ has been showing time-variability.
As discussed in Section~\ref{discuss}, the variability of $PA$ is observed even when we focus only on the $V$-band measurements presented in Table~\ref{obs_Vband}, and thus the historic $PA$ data are supporting the discussion given in the main text that the $PA$ variability has real physical origin intrinsic to the innermost scattering structure of 3C323.1.

\begin{table*}
	\centering
	\caption{Historic white-light and filtered polarimetric measurements for 3C~323.1. The Galactic ISP is uncorrected.}
	\label{obs_unfilter}
	\begin{tabular}{cclllc}
		\hline
		Date & MJD & Filter & $\hat{p}_{\text{uncorr}}$[\%] & $PA_{\text{uncorr}}$[deg] & Reference\\
		\hline
1977-09-11 & 43397 &None (Steward)& 1.63$\pm$0.24 & 16$\pm$4 & \cite{sto84} \\
1978-05-05 & 43633 &None (UAO)&     2.01$\pm$1.25 & ---     & \cite{sto84} \\
1979-04-02 & 43965 &None (Steward)& 1.48$\pm$0.30 & 19$\pm$6 & \cite{sto84} \\
1979-05-22 & 44015 &None (Steward)& 1.01$\pm$0.20 &  4$\pm$5 & \cite{sto84} \\
1979-05-22 & 44015 &CuSO${}_4$ + B-390 (Steward)  & 0.78$\pm$0.24 & $-$8$\pm$8 & \cite{sto84} \\
1979-05-22 & 44015 &RG715 (Steward) & 0.55$\pm$0.26 & $-$4$\pm$12 & \cite{sto84} \\
1980-03-15 & 44313 &None (Steward)  & 1.42$\pm$0.30 &   8$\pm$6 & \cite{sto84} \\
1990-02-26 & 47948 &None (McDonald) & 1.02$\pm$0.13 & 28.3$\pm$3.6 &\cite{wil11}\\
		\hline
	\end{tabular}
\end{table*}

\begin{figure}
\center{
\includegraphics[clip, width=3.3in]{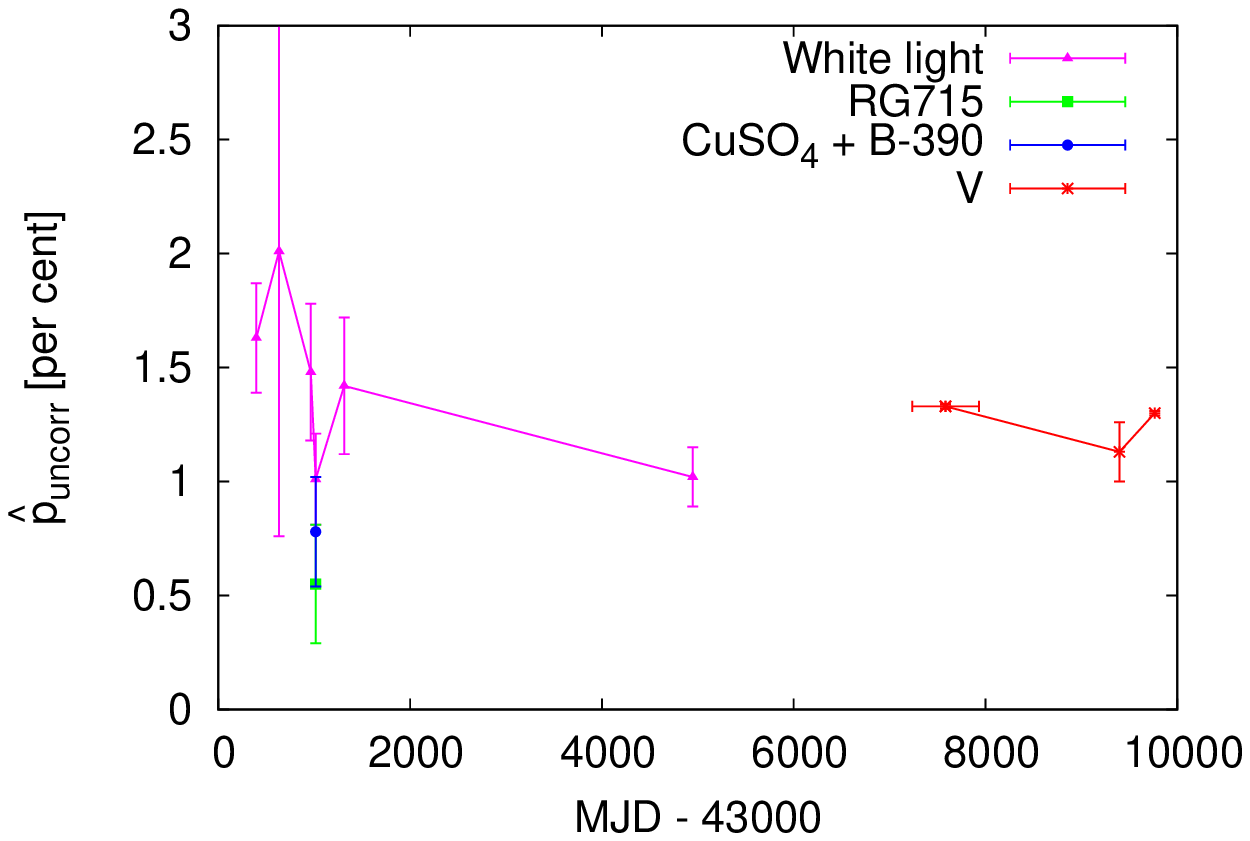}
\includegraphics[clip, width=3.3in]{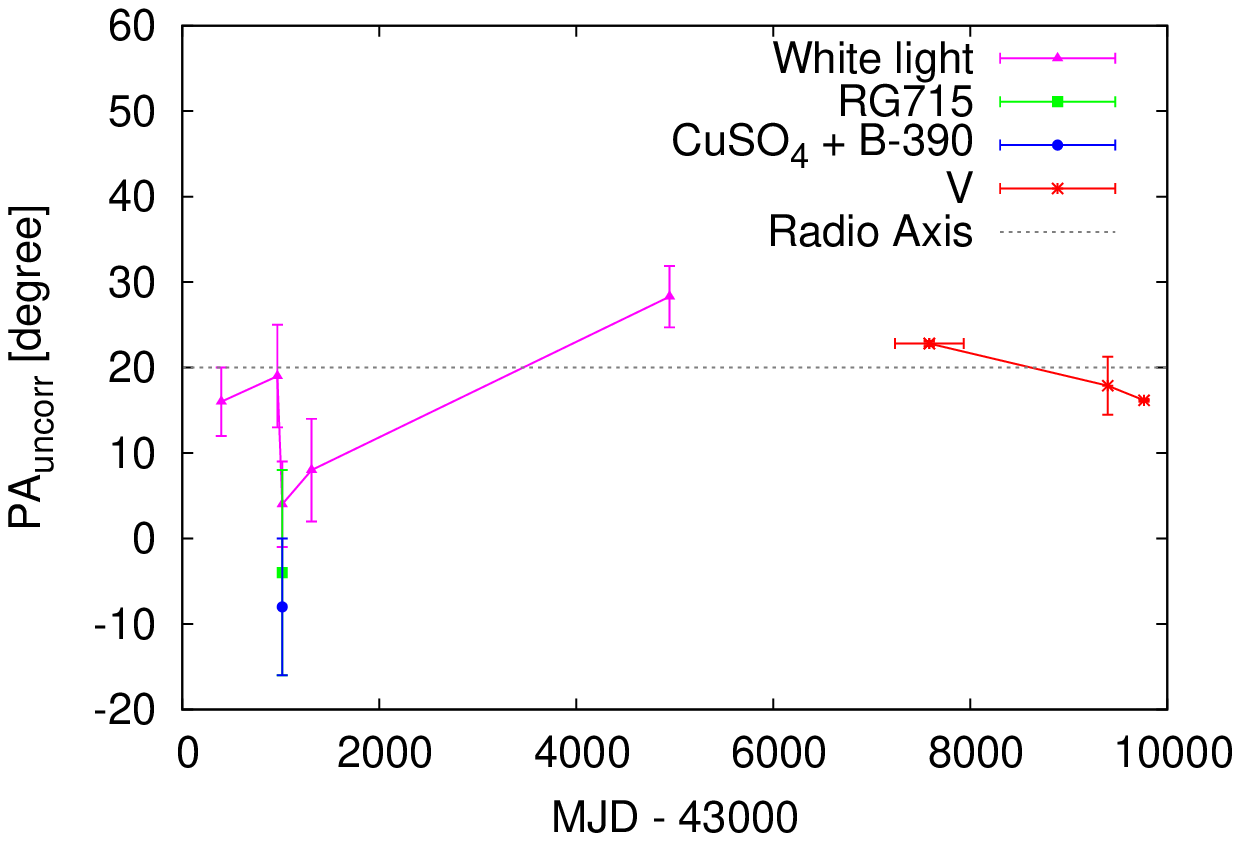}
}
 \caption{
 The time variation of the ISP-uncorrected $\hat{p}$ and $PA$ of 3C~323.1 in 1977-2003.
 3C~323.1's radio jet axis of $20$ deg \citep{kis04} is indicated as a dotted line in the bottom panel.
 }
 \label{fig:summary_pol_nonfilter}
\end{figure}

\bsp	
\label{lastpage}
\end{document}